\documentclass[twocolumn,twocolappendix,tighten,times]{aastex701}

\usepackage{amsmath}

\begin{document}

\title{A semi-analytic model of the bouncing barrier for protoplanetary dust aggregates}

\author[orcid=0000-0003-0947-9962, sname=Arakawa, gname=Sota]{Sota Arakawa}
\affiliation{Center for Mathematical Science and Advanced Technology, Japan Agency for Marine-Earth Science and Technology, 3173-25 Showa-machi, Kanazawa-ku, Yokohama, 236-0001, Japan}
\email[show]{arakawas@jamstec.go.jp}

\author[orcid=0009-0004-3868-373X, sname=Oshiro, gname=Haruto]{Haruto Oshiro}
\affiliation{Department of Earth and Planetary Sciences, Institute of Science Tokyo, 2-12-1 Ookayama, Meguro, Tokyo, 152-8550, Japan}
\email{oshiro.h.aa@m.titech.ac.jp }

\author[orcid=0000-0002-2631-7095, sname=Yoshida, gname=Yuki]{Yuki Yoshida}
\affiliation{Astronomical Institute, Tohoku University, 6-3 Aramaki, Aoba-ku, Sendai, 980-8578, Japan}
\email{yuki.yoshida@astr.tohoku.ac.jp}

\author[orcid=0000-0003-4445-1447, sname=Yoshii, gname=Kiwamu]{Kiwamu Yoshii}
\affiliation{Department of Applied Physics, Tokyo University of Science, 6-3-1, Niijuku, Katsushika, Tokyo, 125-8585, Japan}
\email{qyoshii@rs.tus.ac.jp}


\begin{abstract}

Collisional bouncing limits the growth of dust aggregates in protoplanetary disks, but its dependence on aggregate size, collision velocity, and filling factor remains poorly understood.
Here we develop a semi-analytic model for the sticking probability of colliding dust aggregates.
We divide each aggregate collision into two phases: a compression phase and a separation phase.
The compression phase is described with an elastoplastic contact model, which determines the maximum contact radius and repulsive energy after compression.
The separation phase is treated as fracture of a stochastic network of interparticle bonds, whose fracture energy is evaluated using weakest-link statistics.
The model naturally predicts that larger aggregates bounce more readily because larger contact regions are more likely to contain weak bonds.
Comparison with distinct element method simulations shows that the model reproduces the simulated sticking--bouncing boundary.
Furthermore, applying the calibrated model to moderately porous aggregates inferred from ALMA observations of protoplanetary disks, we find that the predicted bouncing barrier passes through the observationally inferred size--velocity range.
Thus, our semi-analytic model provides a useful framework for predicting the collisional evolution of protoplanetary dust aggregates.

\end{abstract}



\section{Introduction}

Coagulation of submicron-sized dust grains in protoplanetary disks is the first step toward planet formation.
Dust grains form porous aggregates through low-velocity collisions, and the subsequent collisional evolution of these aggregates controls the pathway toward planetesimal formation \citep[e.g.,][]{2012ApJ...752..106O, 2021ApJ...922...16K}.
Thus, understanding aggregate--aggregate collisions is essential for modeling dust growth and interpreting observations of protoplanetary disks \citep[e.g.,][]{2008ARA&A..46...21B, 2014prpl.conf..547J, 2024ARA&A..62..157B}.

One of the key obstacles to dust growth is the bouncing barrier \citep[e.g.,][]{2010A&A...513A..57Z, 2024A&A...682A.144D}.
Laboratory experiments \citep[e.g.,][]{2008ApJ...675..764L, 2010A&A...513A..56G, 2022MNRAS.509.5641S} and numerical simulations \citep[e.g.,][]{2011ApJ...737...36W, 2013A&A...551A..65S} have shown that low-velocity collisions of compact aggregates can result in bouncing rather than sticking.
However, the conditions for sticking and bouncing remain uncertain.
In particular, previous numerical simulations generally used aggregates much smaller than those used in laboratory experiments, and this difference in aggregate size may be one reason for the apparent discrepancy between numerical and experimental sticking--bouncing thresholds \citep{2023ApJ...951L..16A}.

Recent distinct element method (DEM) simulations have clarified several important aspects of this problem.
\citet{2023ApJ...951L..16A} showed that the sticking probability decreases with increasing aggregate radius for aggregates with a fixed filling factor, suggesting that the bouncing barrier is size dependent and should be described statistically.
\citet{2025ApJ...983...75O} also performed collision simulations of moderately compact aggregates and found that the bouncing threshold depends on impact velocity and filling factor.
Their energy analysis further showed that most of the initial impact energy is dissipated during the compression phase of the collision, and that a substantial fraction of the remaining energy is dissipated during the subsequent stretching phase.
These results indicate that the sticking probability can be modeled as a function of aggregate size, collision velocity, and filling factor.

The mechanics of aggregate--aggregate contact formation has also been examined through DEM simulations combined with analyses based on effective continuum descriptions.
\citet{2024arXiv240815573A} showed that the maximum compression length of colliding aggregates is proportional to the aggregate radius and increases with collision velocity.
This implies that the radius of the aggregate--aggregate contact area is also proportional to the aggregate radius and increases with collision velocity \citep{2025ApJ...995..207T}.
\citet{2024arXiv240815573A} demonstrated that these size and velocity dependences are well reproduced by a contact model for elastoplastic spheres originally proposed by \citet{doi:10.1080/14786443008565033}.
Thus, the compression phase of aggregate collisions can be approximately described by macroscopic elastoplastic contact mechanics.
However, the decompression phase after maximum compression deviates from the Andrews' model: the interaggregate interaction becomes attractive near the contact region, and separation requires breaking the interaggregate bonds formed during collision.

In this study, we develop a semi-analytic model for the sticking probability of low-velocity collisions between dust aggregates.
We first use an elastoplastic contact model of \citet{doi:10.1080/14786443008565033} to calculate the maximum contact radius $a_{\rm max}$ and the repulsive energy $K_{\rm rep}$ available after maximum compression (Section \ref{sec:macro}).
We then model the contact region as a stochastic network of interparticle bonds and evaluate its fracture energy $E_{\rm break}$ using weakest-link statistics \citep[e.g.,][]{Weibull1939, 2020MNRAS.496.1667K}, as described in Section \ref{sec:micro}.
Finally, the sticking probability is obtained by comparing $K_{\rm rep}$ with the fracture energy $E_{\rm break}$ required to separate the contact region.

The central idea of our model is that a larger contact region contains more interparticle bonds and is therefore more likely to include statistically weak bonds.
If the bond-breaking energies are heterogeneous, the effective fracture energy of the contact region increases more slowly than the aggregate mass.
As a result, larger aggregates can more readily break their contact region during separation, providing a natural explanation for why the sticking probability decreases with aggregate size.

\section{Model}
\label{sec:model}

We consider a head-on collision between two equal-mass aggregates of radius $R_{\rm agg}$, volume filling factor $\phi$, and radius of constituent grains $r_{1}$.
The material density of the constituent grains is denoted by $\rho_{\rm mat}$, so that the bulk density of the aggregate is $\rho_{\rm agg} = \phi \rho_{\rm mat}$.
The mass of one aggregate is $m_{\rm agg} = {( 4 \pi / 3 )} \rho_{\rm agg} {R_{\rm agg}}^{3}$.
For a collision with relative velocity $v$, the initial kinetic energy (in the center-of-mass frame) is
\begin{equation}
K_{\rm ini} = \frac{1}{4} m_{\rm agg} v^{2} = \frac{8 \pi}{3} \rho_{\rm agg} v^{2} {R^{*}}^{3},
\label{eq:K_ini}
\end{equation}
where $R^{*} = R_{\rm agg} / 2$ is the reduced radius.

We divide the collision into two phases. 
The first phase is the compression phase from first contact to maximum compression (Section \ref{sec:macro}).
In this phase, we treat the aggregates as effective elastoplastic spheres and calculate the maximum compression length $\delta_{\rm max}$, the maximum contact radius $a_{\rm max}$, and the repulsive energy $K_{\rm rep}$ available after maximum compression \citep[see also][]{2024arXiv240815573A}.
The second phase is the separation phase, in which the contact region is stretched and fractured (Section \ref{sec:micro}).
In this phase, we model the contact region as a network of interparticle bonds and statistically evaluate its fracture energy $E_{\rm break}$.
The final sticking or bouncing outcome is determined by comparing the repulsive energy $K_{\rm rep}$ with the fracture energy of the contact region, $E_{\rm break}$.

\subsection{Macroscopic Model of Contact Formation}
\label{sec:macro}

We regard the aggregates as effective elastoplastic spheres with reduced Young's modulus $E^{*}$ and yield stress $\sigma_{\rm y}$ (see Appendix \ref{app:yield} for details).
The critical compression length for yielding is given by \citep{doi:10.1080/14786443008565033}
\begin{equation}
\delta_{\rm crit} = {\left( \frac{\pi \sigma_{\rm y}}{2 E^{*}} \right)}^{2} R^{*},
\label{eq:delta_crit}
\end{equation}
and the corresponding critical collision velocity $v_{\rm crit}$ is
\begin{equation}
v_{\rm crit} = {\left( \frac{E^{*}}{5 \pi \rho_{\rm agg}} \right)}^{1/2} {\left( \frac{\pi \sigma_{\rm y}}{2 E^{*}} \right)}^{5/2}.
\label{eq:v_crit}
\end{equation}
We also define the critical kinetic energy $K_{\rm crit}$ as the value of $K_{\rm ini}$ at $v = v_{\rm crit}$, and $K_{\rm crit}$ is
\begin{equation}
K_{\rm crit} = \frac{8}{15} {\left( \frac{\pi \sigma_{\rm y}}{2 E^{*}} \right)}^{5} E^{*} {R^{*}}^{3}.
\end{equation}

For $v \le v_{\rm crit}$, the collision remains in the elastic regime, and the maximum compression length is obtained from the following equation \citep{hertz1896miscellaneous}:
\begin{equation}
\frac{\delta_{\rm max}}{\delta_{\rm crit}} = {\left( \frac{v}{v_{\rm crit}} \right)}^{4/5}.
\label{eq:delta_max_1}
\end{equation}
In contrast, for $v > v_{\rm crit}$, plastic deformation occurs during compression, and $\delta_{\rm max}$ is obtained by solving the quadratic equation \citep{doi:10.1080/14786443008565033}:
\begin{equation}
\frac{15}{8} {\left[ \frac{1}{2} {\left( \frac{\delta_{\rm max}}{\delta_{\rm crit}} \right)}^{2} - \frac{1}{3} \frac{\delta_{\rm max}}{\delta_{\rm crit}} + \frac{1}{10}  \right]} = {\left( \frac{v}{v_{\rm crit}} \right)}^{2}.
\label{eq:delta_max_2}
\end{equation}

The maximum contact radius $a_{\rm max}$ is given by
\begin{equation}
a_{\rm max} = \sqrt{R^{*} \delta_{\rm max}}.
\label{eq:a_max}
\end{equation}
Equation \eqref{eq:delta_crit} gives $\delta_{\rm crit} \propto R^{*}$, and for fixed collision velocity and fixed aggregate mechanical properties ($E^{*}$ and $\sigma_{\rm y}$), Equations \eqref{eq:delta_max_1} and \eqref{eq:delta_max_2} show that $\delta_{\rm max} \propto R^{*}$, and therefore Equation \eqref{eq:a_max} gives $a_{\rm max} \propto R^{*}$.
Thus, the geometry of the contact region scales self-similarly with aggregate size.
These scaling relations have been confirmed in DEM simulations by \citet{2024arXiv240815573A} and \citet{2025ApJ...995..207T}.

The elastoplastic contact model of \citet{doi:10.1080/14786443008565033} also provides an expression for the repulsive energy $K_{\rm rep}$ available after maximum compression.
For $v \le v_{\rm crit}$, the collision remains in the elastic regime, and
\begin{equation}
\frac{K_{\rm rep}}{K_{\rm crit}} = \frac{K_{\rm ini}}{K_{\rm crit}} ={\left( \frac{v}{v_{\rm crit}} \right)}^{2}.
\end{equation}
In contrast, for $v > v_{\rm crit}$, plastic deformation causes energy dissipation, and $K_{\rm rep}$ is given by
\begin{equation}
\frac{K_{\rm rep}}{K_{\rm crit}} = \frac{1}{3} {\left[ \sqrt{ 30 {\left( \frac{v}{v_{\rm crit}} \right)}^{2} - 5 } - 2 \right]}.
\label{eq:K_rep_2}
\end{equation}

\subsection{Microscopic Model of Fracture}
\label{sec:micro}

At maximum compression, the two aggregates form a contact region of radius $a_{\rm max}$.
This region contains many interparticle bonds, which must be broken for the two aggregates to separate.

\begin{figure*}[]
\centering
\includegraphics[width = 0.8\textwidth]{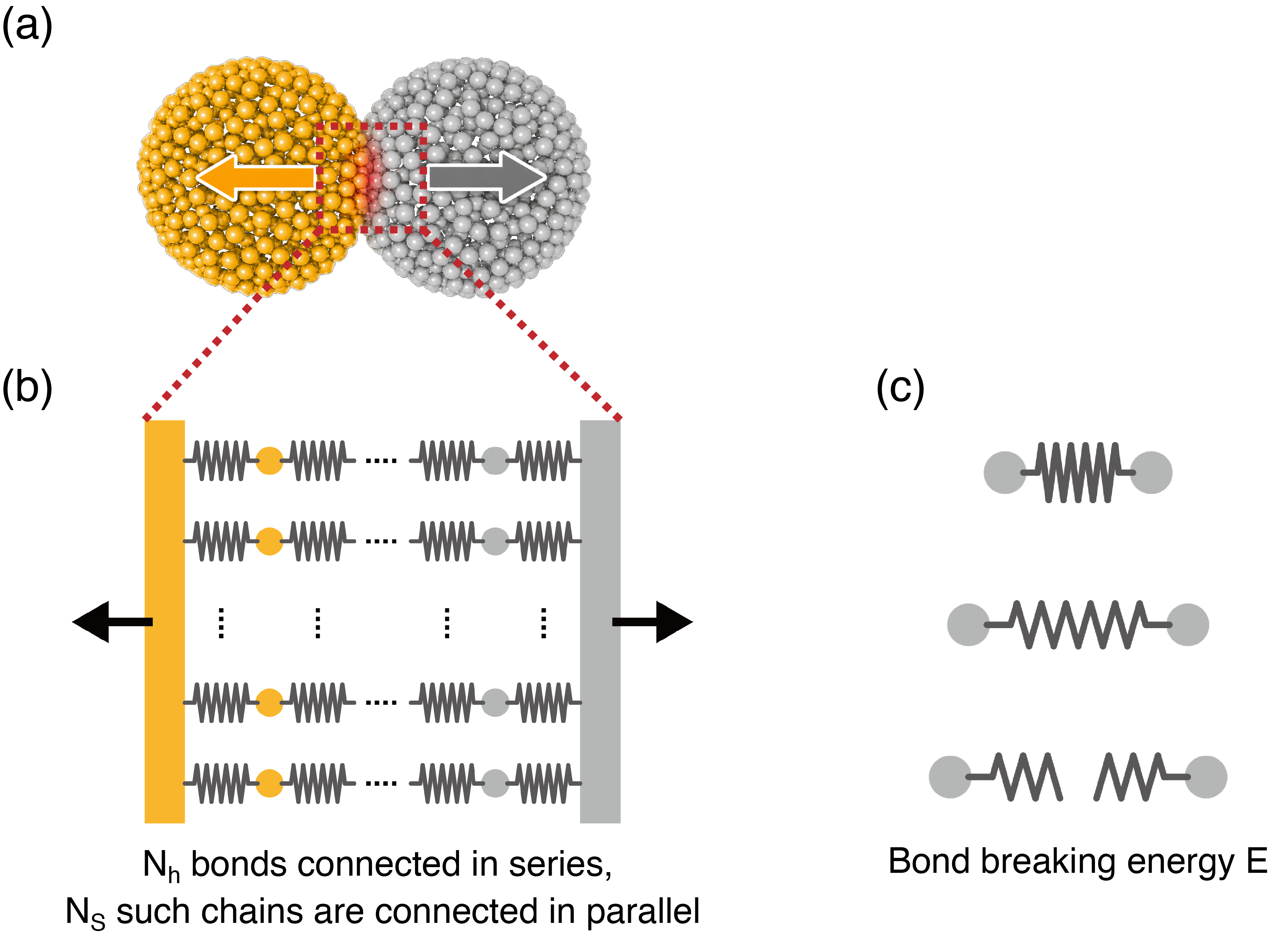}
\caption{
Schematic illustrations of the fracture model for the aggregate--aggregate contact region.
(a) Two aggregates in contact during the separation phase.
Energy dissipation associated with bond breaking occurs in the aggregate--aggregate contact region with radius $a_{\rm max}$.
(b) Bond network sandwiched between two rigid plates.
A single chain consists of $N_{\rm h}$ bonds connected in series, and $N_{\rm S}$ such chains are connected in parallel.
(c) Breakage of an interparticle bond.
Each bond has its own breaking energy $E$, whose probability distribution is assumed to be given by Equation \eqref{eq:P_1}.
}
\label{fig:schematic}
\end{figure*}

We idealize the contact region as a bond network sandwiched between two rigid plates representing the two aggregates during separation (Figure \ref{fig:schematic}).
A single chain consists of $N_{\rm h}$ bonds connected in series, and $N_{\rm S}$ such chains are connected in parallel.
The number of bonds along the thickness direction of the contact region is estimated as
\begin{equation}
N_{\rm h} \simeq C_{\rm h} \frac{a_{\rm max}}{r_{1}},
\label{eq:N_h}
\end{equation}
where $C_{\rm h}$ is a geometrical factor of order unity.
The number of independent chains in the contact area is estimated as
\begin{equation}
N_{\rm S} \simeq C_{\rm S} \frac{Z}{2} \phi {\left( \frac{a_{\rm max}}{r_{1}} \right)}^{2},
\label{eq:N_S}
\end{equation}
where $Z$ is the mean coordination number of the aggregate (see Appendix \ref{app:Z}) and $C_{\rm S}$ is another geometrical factor of order unity.
The factor $Z / 2$ accounts for the fact that each contact is shared by two grains.
In this study, we set $C_{\rm h} = C_{\rm S} = 1$ for simplicity.

\subsubsection{Distribution of Bond-breaking Energies}

We assume that each interparticle bond has its own breaking energy $E$. 
The bond-breaking energy may vary because of local differences in coordination number, bond orientation, deformation history, and the relative importance of normal and tangential motions.

We characterize this heterogeneity using the cumulative probability $P {\left( E < E' \right)}$, defined as the probability that the breaking energy $E$ of a bond is smaller than $E'$.
We assume that this probability follows a power law:
\begin{equation}
P {\left( E < E' \right)} = {\left( \frac{E'}{E_{1}} \right)}^{\alpha} \qquad {\left( 0 < E' < E_{1} \right)}.
\label{eq:P_1}
\end{equation}
The breaking energy $E$ is assumed to take values in the range $0 < E < E_{1}$, and the characteristic breaking energy $E_{1}$ is written as
\begin{equation}
E_{1} = E_{\rm b} + \beta E_{\rm r},
\label{eq:E_1}
\end{equation}
where $E_{\rm b}$ is the energy required to break the normal adhesive bond between two grains, $E_{\rm r}$ is the energy dissipated when one grain rolls by $90^{\circ}$ over another, and $\beta$ is a dimensionless weighting parameter (see Appendix \ref{app:material} for details).

The exponent $\alpha$ in Equation \eqref{eq:P_1} controls the low-energy tail of the bond-breaking-energy distribution.
A larger $\alpha$ corresponds to a more homogeneous bond population, whereas a smaller $\alpha$ corresponds to a population with a larger fraction of weak bonds.

\subsubsection{Weakest-link Statistics}
\label{sec:weakest}

A chain consisting of $N_{\rm h}$ bonds in series breaks when its weakest bond breaks.
Let $E_{\rm min}$ be the minimum breaking energy among the $N_{\rm h}$ bonds.
The probability that $E_{\rm min} < E'$ is
\begin{eqnarray}
P {\left( E_{\rm min} < E' \right)} & = & 1 - {\left[ 1 - P {\left( E < E' \right)} \right]}^{N_{\rm h}} \nonumber \\
                                    & = & 1 - {\left[ 1 - {\left( \frac{E'}{E_{1}} \right)}^{\alpha} \right]}^{N_{\rm h}}.
\end{eqnarray}

When the chain is stretched, we assume that comparable amounts of energy are dissipated in individual bonds.
Therefore, when the weakest bond breaks, the total energy dissipated in the chain is
\begin{equation}
E_{\rm chain} \simeq N_{\rm h} E_{\rm min}.
\end{equation}
The DEM simulations of \citet{2025ApJ...983...75O} show that energy dissipation during the separation phase occurs within spherical regions with radii comparable to $a_{\rm max}$ (see their Figure 10), supporting the qualitative validity of our assumption.
In the limit $N_{\rm h} \gg 1$, the mean fracture energy of a chain is
\begin{eqnarray}
{\left\langle E_{\rm chain} \right\rangle} & =      & \int_{0}^{E_{1}}~N_{\rm h} E' \frac{{\rm d}P {\left( E_{\rm min} < E' \right)}}{{\rm d}E'}~{\rm d}E' \nonumber \\
                                           & \simeq & \Gamma{\left( 1 + \frac{1}{\alpha} \right)} E_{1} {N_{\rm h}}^{1 - 1 / \alpha},
\label{eq:E_chain}
\end{eqnarray}
and its variance is
\begin{eqnarray}
{\rm Var}{\left( E_{\rm chain} \right)} \simeq & & {\left[ \Gamma{\left( 1 + \frac{2}{\alpha} \right)} - \Gamma{\left( 1 + \frac{1}{\alpha} \right)}^{2} \right]} \times \nonumber \\
                                               & & {E_{1}}^2 {N_{\rm h}}^{2 {( 1 - 1 / \alpha )}}.
\end{eqnarray}

The important point is that the chain fracture energy does not scale linearly with $N_{\rm h}$ when the bond strengths are heterogeneous.
A longer chain is more likely to contain a weak bond, and this weakest-link effect leads to ${\left\langle E_{\rm chain} \right\rangle} \propto {N_{\rm h}}^{1 - 1 / \alpha}$.

The entire contact region is modeled as $N_{\rm S}$ chains connected in parallel.
Thus, the total fracture energy is
\begin{equation}
E_{\rm break} = \sum_{i = 1}^{N_{\rm S}} E_{{\rm chain}, i}.
\end{equation}
Assuming that the chain fracture energies are independent, the mean value is
\begin{equation}
{\left\langle E_{\rm break} \right\rangle} = {\left\langle E_{\rm chain} \right\rangle} \times N_{\rm S},
\label{eq:E_break}
\end{equation}
and the variance is
\begin{equation}
{\rm Var}{\left( E_{\rm break} \right)} = {\rm Var}{\left( E_{\rm chain} \right)} \times N_{\rm S}.
\end{equation}
For large $N_{\rm S}$, the central limit theorem allows the distribution of $E_{\rm break}$ to be approximated by a normal distribution with mean ${\left\langle E_{\rm break} \right\rangle}$ and variance ${\rm Var}{\left( E_{\rm break} \right)}$.

\subsection{Criterion for Sticking and Bouncing}
\label{sec:f_stick}

The two aggregates can separate only if the repulsive energy exceeds the fracture energy of the contact region: $K_{\rm rep} > E_{\rm break}$.
In this case, the contact region is broken and the collision results in bouncing.
If $K_{\rm rep} \le E_{\rm break}$, the contact region cannot be fully broken and the two aggregates remain stuck.
Therefore, the sticking probability is
\begin{eqnarray}
f_{\rm stick} & =      & P{\left( K_{\rm rep} \le E_{\rm break} \right)} \nonumber \\
              & \simeq & \frac{1}{2} {\rm erfc} {\left[ \frac{K_{\rm rep} - {\left\langle E_{\rm break} \right\rangle}}{\sqrt{2 {\rm Var}{\left( E_{\rm break} \right)}}} \right]},
\label{eq:f_stick}
\end{eqnarray}
where ${\rm erfc}$ is the complementary error function.
It follows directly from Equation \eqref{eq:f_stick} that the condition $f_{\rm stick} = 50\%$ corresponds to ${\left\langle E_{\rm break} \right\rangle} / K_{\rm rep} = 1$.

\section{Size and Velocity Dependence}
\label{sec:dependence}

Equation \eqref{eq:f_stick} indicates that the sticking probability $f_{\rm stick}$ can be understood in terms of the competition between the repulsive energy $K_{\rm rep}$ and the mean fracture energy ${\left\langle E_{\rm break} \right\rangle}$.
In this section, we derive the size and velocity dependences of the ratio ${\left\langle E_{\rm break} \right\rangle} / K_{\rm rep}$.

\subsection{Origin of the Size Dependence}

The present model naturally explains the size dependence of the sticking probability reported by \citet{2023ApJ...951L..16A} and \citet{2025ApJ...983...75O}.
For fixed collision velocity and filling factor, the elastoplastic contact model gives $a_{\rm max} \propto R_{\rm agg}$, which implies $N_{\rm h} \propto R_{\rm agg}$ and $N_{\rm S} \propto {R_{\rm agg}}^{2}$.
The mean fracture energy of the contact region scales as ${\left\langle E_{\rm break} \right\rangle} \propto N_{\rm S} N_{\rm h}^{1 - 1 / \alpha}$ (Equations \eqref{eq:E_chain} and \eqref{eq:E_break}).
Thus,
\begin{equation}
{\left\langle E_{\rm break} \right\rangle} \propto {R_{\rm agg}}^{3 - 1 / \alpha}.
\end{equation}

For fixed collision velocity, $K_{\rm rep}$ scales with aggregate mass and hence
\begin{equation}
K_{\rm rep} \propto {R_{\rm agg}}^{3}.
\end{equation}
Therefore,
\begin{equation}
\frac{ {\left\langle E_{\rm break} \right\rangle} }{K_{\rm rep}} \propto {R_{\rm agg}}^{- 1 / \alpha}.
\label{eq:ratio_R_agg}
\end{equation}
This means that the fracture energy required to break the contact region becomes smaller relative to the repulsive energy for larger aggregates. 
Consequently, larger aggregates are more likely to bounce.

\subsection{Dependence on Collision Velocity}

We next consider how the sticking probability depends on the collision velocity for fixed aggregate radius and filling factor.
In the present model, the collision velocity affects both the repulsive energy $K_{\rm rep}$ and the fracture energy $E_{\rm break}$ of the contact region.

For $v \le v_{\rm crit}$, the collision remains in the elastic regime.
In this regime, the repulsive energy is equal to the initial kinetic energy, and $K_{\rm rep} \propto v^{2}$.
The maximum compression length scales as $\delta_{\rm max} \propto v^{4/5}$ (Equation \eqref{eq:delta_max_1}), and therefore the maximum contact radius scales as
\begin{equation}
a_{\rm max} = \sqrt{R^{*} \delta_{\rm max}} \propto v^{2/5}.
\end{equation}
Because $N_{\rm h} \propto a_{\rm max}$ and $N_{\rm S} \propto a_{\rm max}^{2}$, the mean fracture energy of the contact region scales as
\begin{equation}
{\left\langle E_{\rm break} \right\rangle} \propto N_{\rm S} N_{\rm h}^{1 - 1 / \alpha} \propto
v^{6/5 - 2 / {\left( 5 \alpha \right)}} .
\end{equation}
Thus,
\begin{equation}
\frac{ {\left\langle E_{\rm break} \right\rangle} }{K_{\rm rep}} \propto v^{- 4/5 - 2 / {\left( 5 \alpha \right)}}.
\label{eq:ratio_v_1}
\end{equation}
In the elastic regime, increasing the collision velocity makes the repulsive energy grow faster than the fracture energy.
Therefore, bouncing becomes more likely as the collision velocity increases.

In contrast, for $v > v_{\rm crit}$, plastic deformation occurs during compression.
In this regime, part of the initial kinetic energy is dissipated by plastic deformation, and the repulsive energy grows more slowly than $K_{\rm ini}$.
In the high-velocity limit (i.e., $v \gg v_{\rm crit}$), the repulsive energy scales as $K_{\rm rep} \propto v$ (Equation \eqref{eq:K_rep_2}).
At the same time, the contact region becomes larger with increasing collision velocity.
In the rigid-plastic limit, the maximum compression length scales as $\delta_{\rm max} \propto v$ (Equation \eqref{eq:delta_max_2}), and hence
\begin{equation}
a_{\rm max} \propto v^{1/2}.
\end{equation}
The mean fracture energy then scales as
\begin{equation}
{\left\langle E_{\rm break} \right\rangle} \propto v^{3/2 - 1 / {\left( 2 \alpha \right)}}.
\end{equation}
Therefore,
\begin{equation}
\frac{ {\left\langle E_{\rm break} \right\rangle} }{K_{\rm rep}} \propto v^{1/2 - 1 / {\left( 2 \alpha \right)}}.
\end{equation}

At high velocities, plastic dissipation reduces the fraction of kinetic energy recovered as repulsive energy, while the contact region becomes larger and more difficult to break.
As a result, for $\alpha > 1$, the model predicts that bouncing is favored at intermediate collision velocities rather than at the highest velocities, consistent with the numerical results of \citet{2025ApJ...983...75O}.

\section{Results}
\label{sec:results}

In this section, we compare the sticking probability predicted by our semi-analytic model with the numerical results of \citet{2025ApJ...983...75O} and examine the validity of the model.
We also investigate how the model predictions depend on the parameters $\alpha$ and $\beta$, which characterize the distribution and magnitude of the bond-breaking energies, respectively.

To make a direct comparison with the simulations of \citet{2025ApJ...983...75O}, we assume that the constituent grains are monodisperse icy grains with radius $r_{1} = 100~{\rm nm}$.
We adopt the same material parameters as those used in \citet{2025ApJ...983...75O}.
For the macroscopic mechanical properties of dust aggregates composed of these icy grains, we use the mechanical strength reported by \citet{2023ApJ...953....6T}.
Details of the macroscopic and microscopic material properties adopted in this study are described in Appendices \ref{app:yield} and \ref{app:material}, respectively.

\subsection{Dependence on $\alpha$}
\label{sec:alpha}

\begin{figure*}[]
\centering
\gridline{\fig{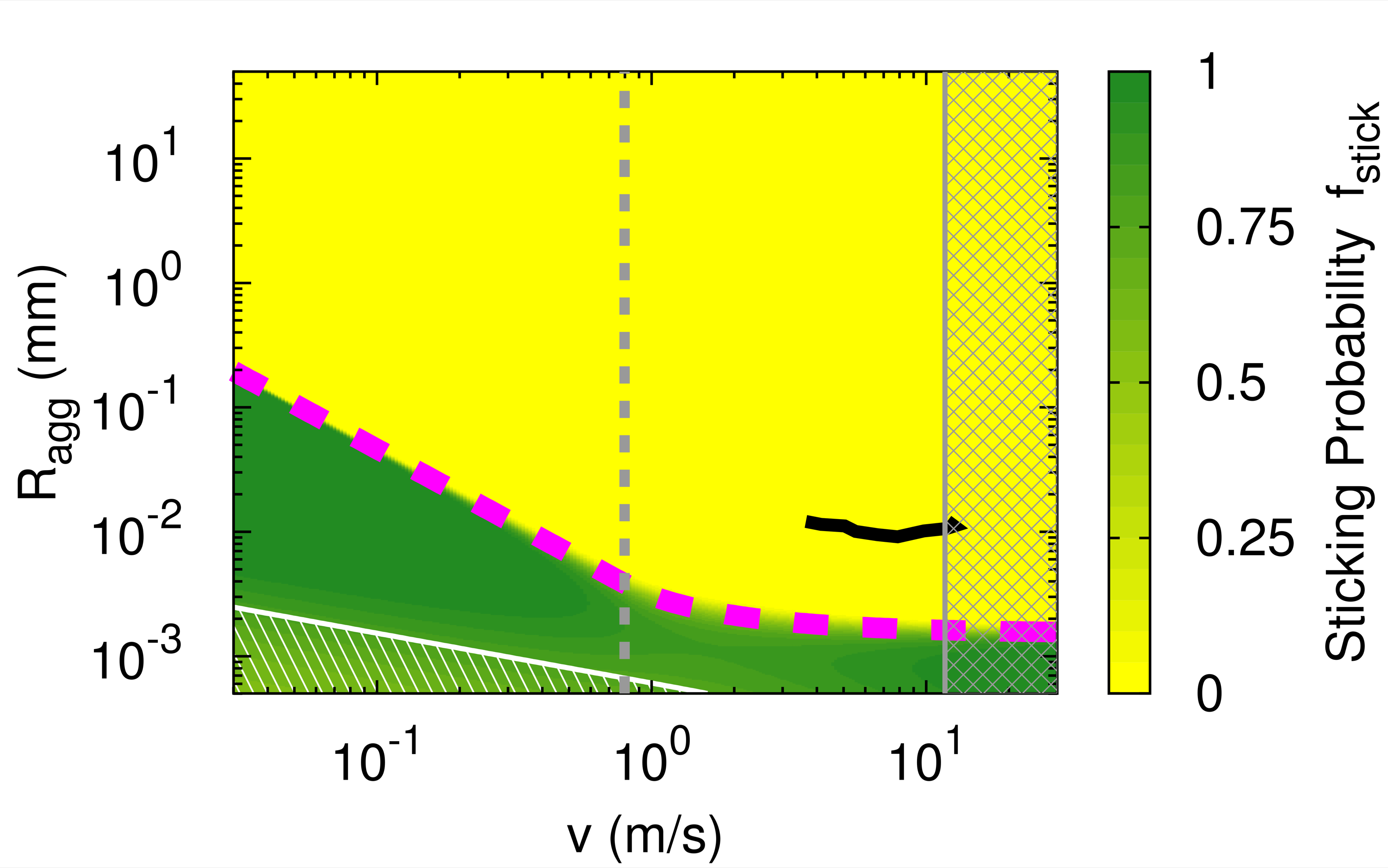}{\columnwidth}{(a) $\phi = 0.4$, $\alpha = 1$, and $\beta = 1$}
          \fig{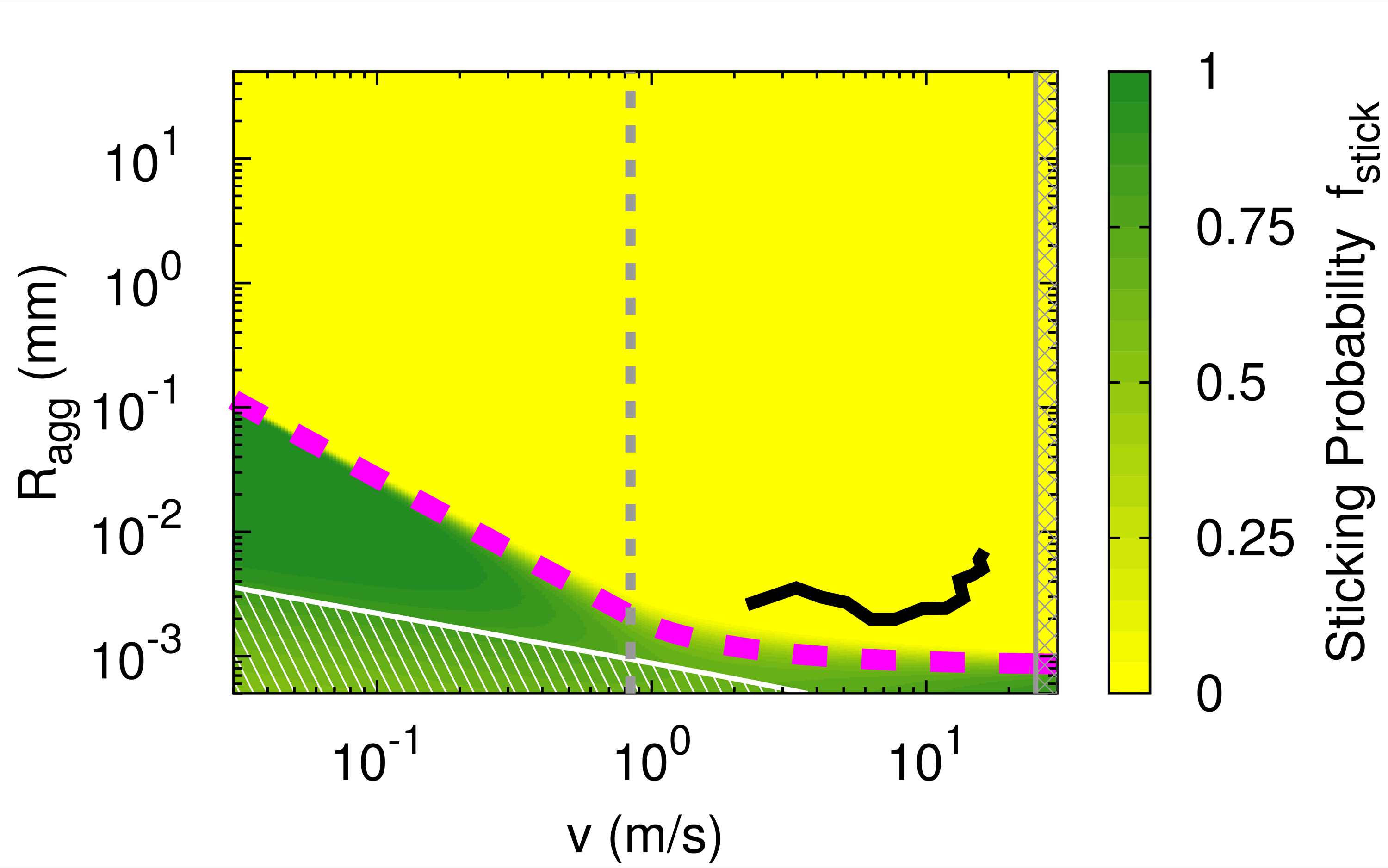}{\columnwidth}{(b) $\phi = 0.5$, $\alpha = 1$, and $\beta = 1$}}
\gridline{\fig{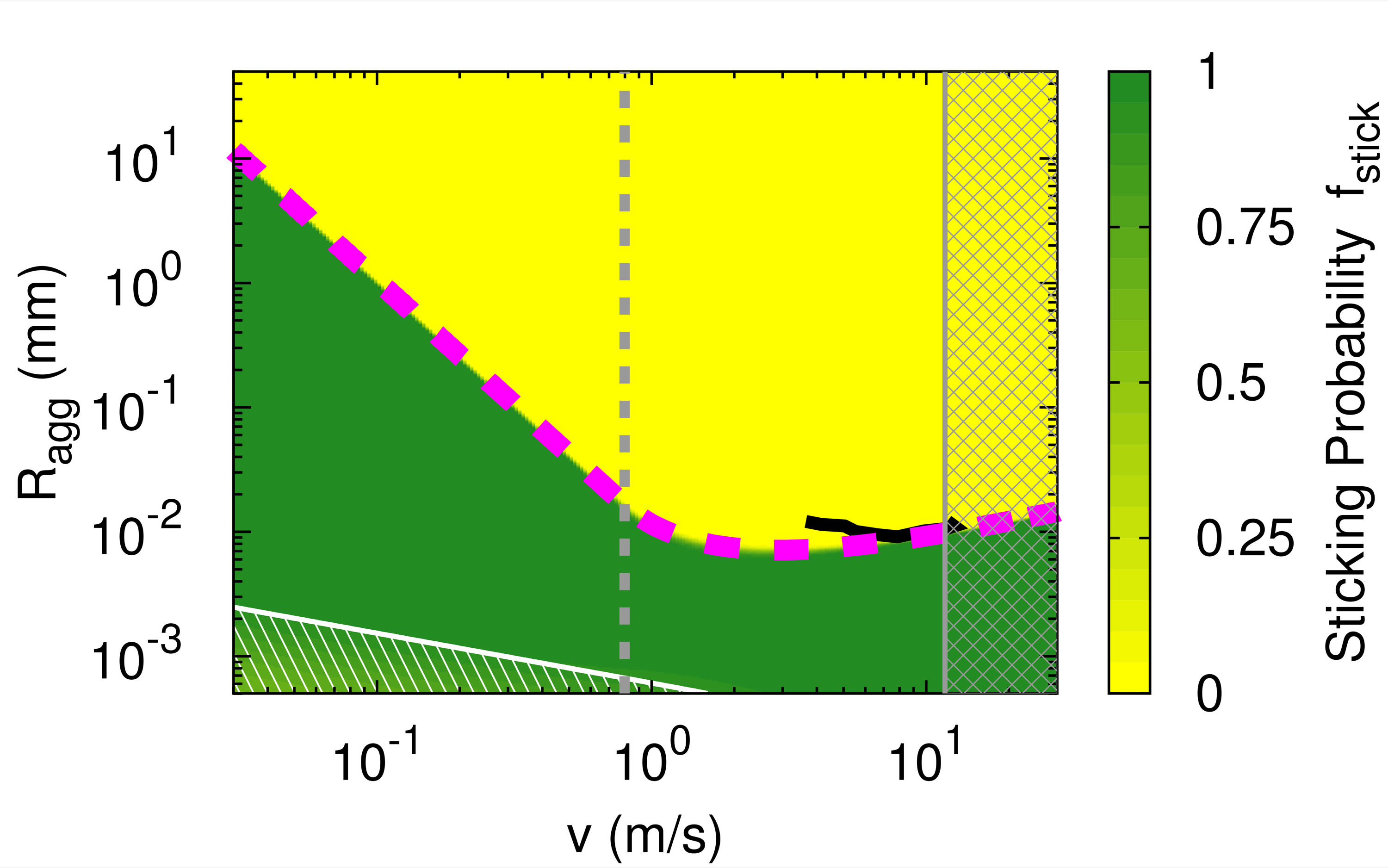}{\columnwidth}{(c) $\phi = 0.4$, $\alpha = 2$, and $\beta = 1$}
          \fig{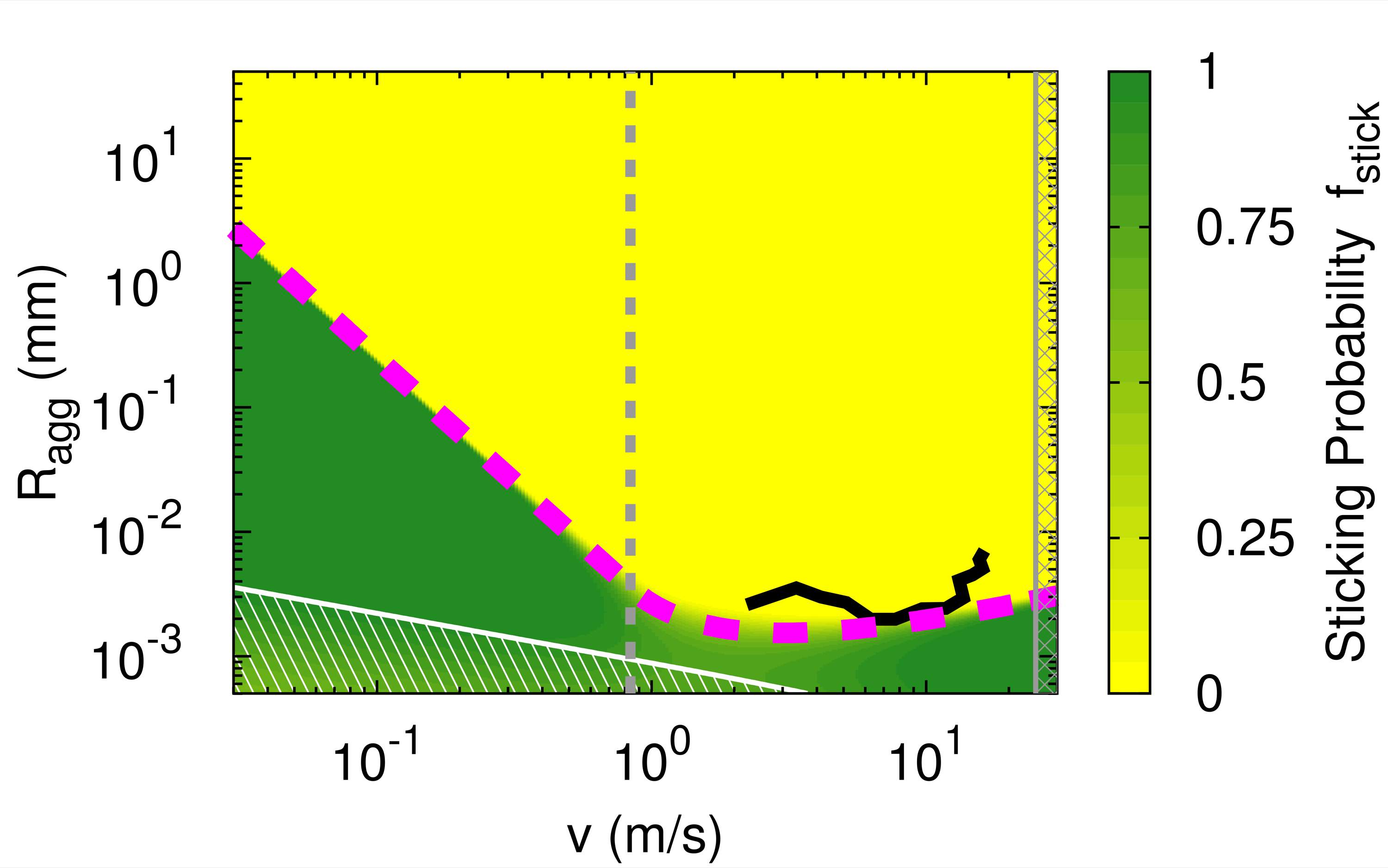}{\columnwidth}{(d) $\phi = 0.5$, $\alpha = 2$, and $\beta = 1$}}
\gridline{\fig{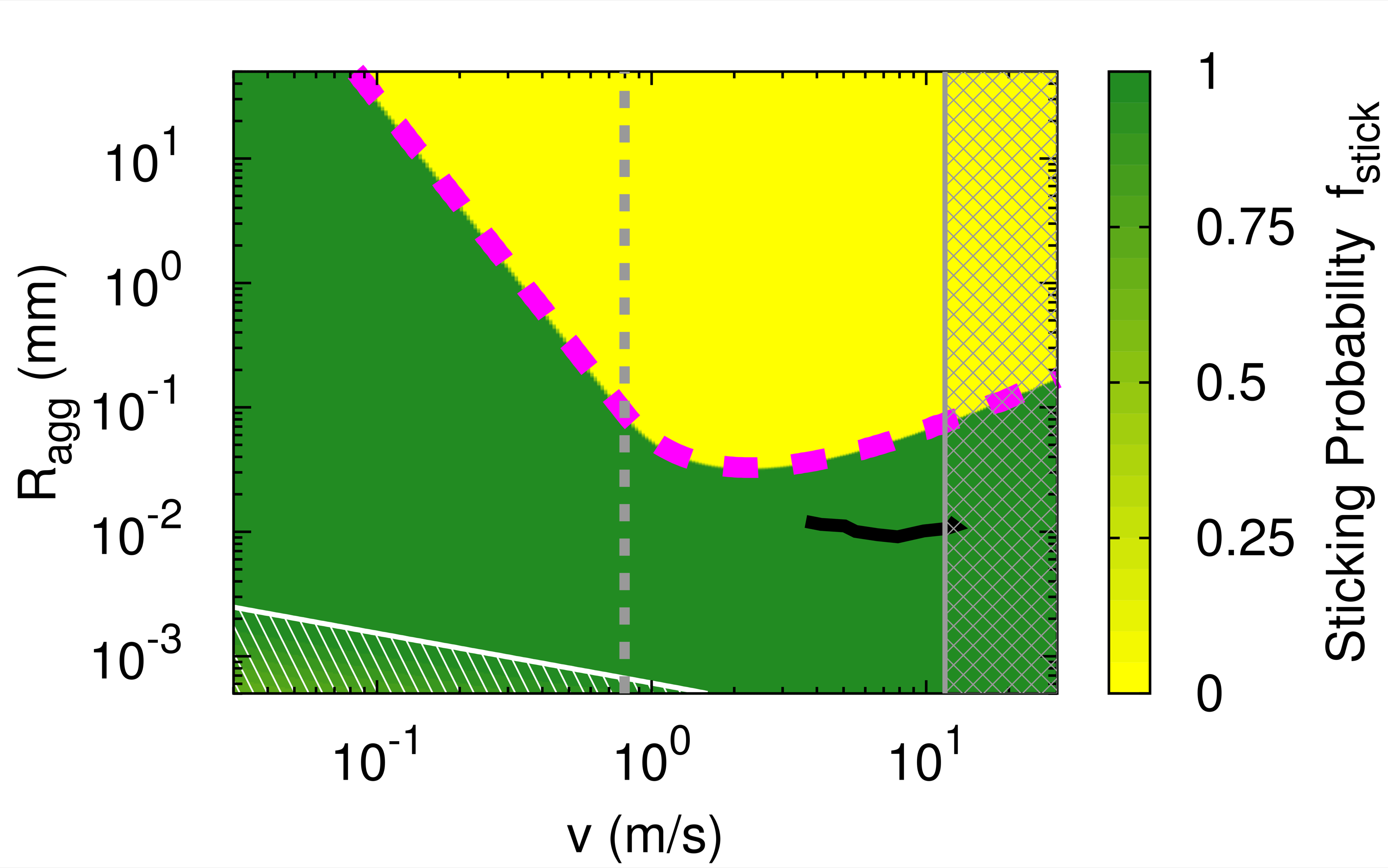}{\columnwidth}{(e) $\phi = 0.4$, $\alpha = 3$, and $\beta = 1$}
          \fig{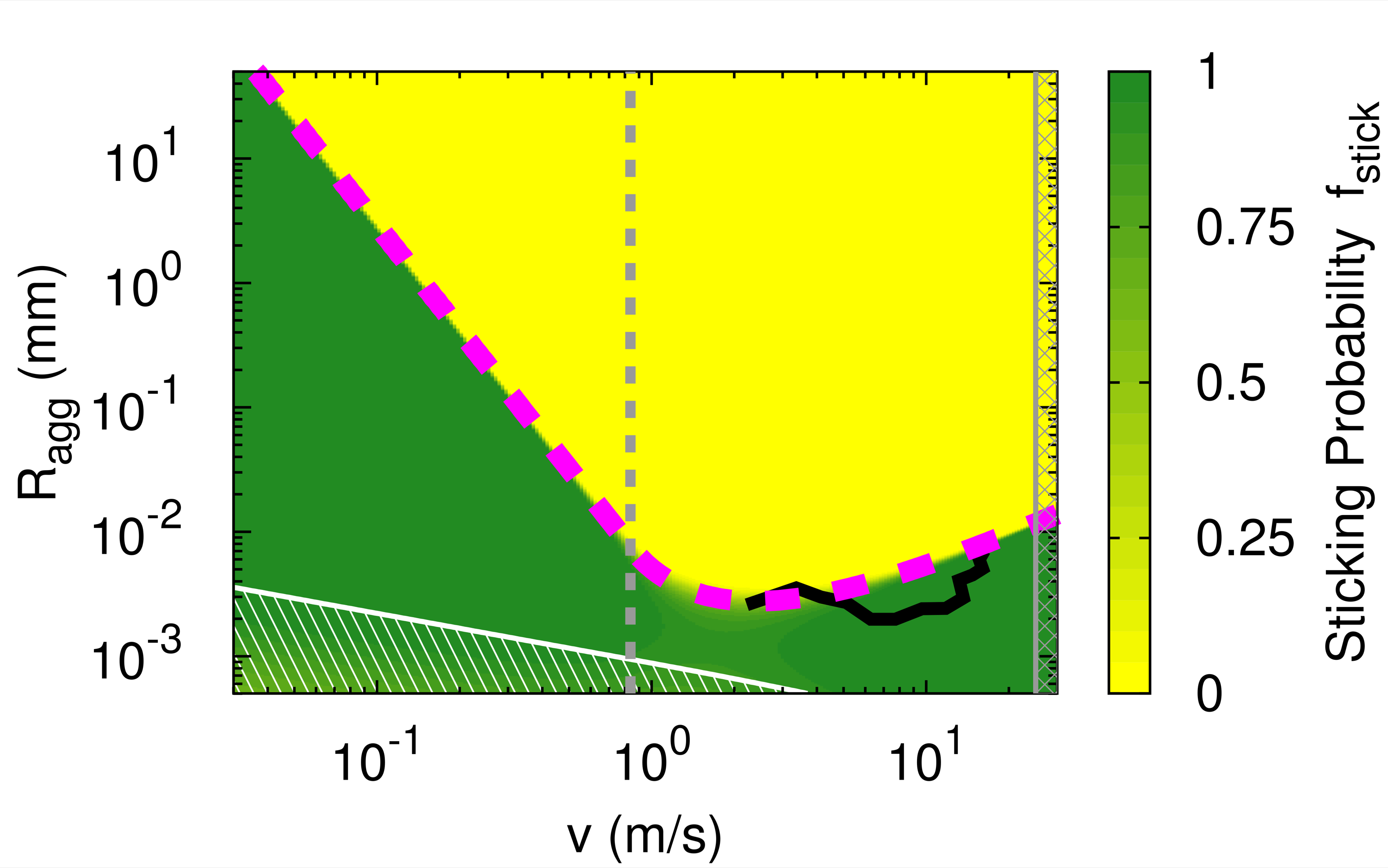}{\columnwidth}{(f) $\phi = 0.5$, $\alpha = 3$, and $\beta = 1$}}
\caption{
Sticking probability $f_{\rm stick}$ in the $v$--$R_{\rm agg}$ plane.
Here we fix $\beta = 1$.
Each panel shows the result for a different set of $\phi$ and $\alpha$.
The color scale represents $f_{\rm stick}$ calculated from Equation \eqref{eq:f_stick}.
The magenta dashed curves denote the $50\%$ sticking condition predicted by our model, $R_{50}$.
The black lines denote the $50\%$ sticking condition obtained from the numerical simulations of \citet{2025ApJ...983...75O}.
The vertical dashed lines indicate the critical velocity $v_{\rm crit}$ for the onset of plastic deformation.
The vertical solid lines indicate the velocity $v_{\rm A}$ at which $\delta_{\rm max} = R^{*}$, and the cross-hatched regions correspond to $v > v_{\rm A}$.
Since the elastoplastic contact model \citep{doi:10.1080/14786443008565033} assumes small deformations of colliding spheres, its applicability is uncertain in these regions.
The white solid lines indicate the contours where $N_{\rm h} = 1$, while the hatched regions denote the parameter space where $N_{\rm h} < 1$.
Our model for estimating ${\left\langle E_{\rm break} \right\rangle}$ is not applicable in these regions.
}
\label{fig:alpha}
\end{figure*}

In Section \ref{sec:alpha}, we examine how the parameter $\alpha$, which characterizes the shape of the bond-breaking-energy distribution, affects the size and velocity dependence of the sticking probability.
Here, we fix $\beta = 1$ and vary $\alpha$.
Figure \ref{fig:alpha} shows the sticking probability $f_{\rm stick}$ in the $v$--$R_{\rm agg}$ plane for $\phi = 0.4$ and $\phi = 0.5$.
The color scale represents $f_{\rm stick}$ calculated from Equation \eqref{eq:f_stick}.

\begin{figure*}[]
\centering
\gridline{\fig{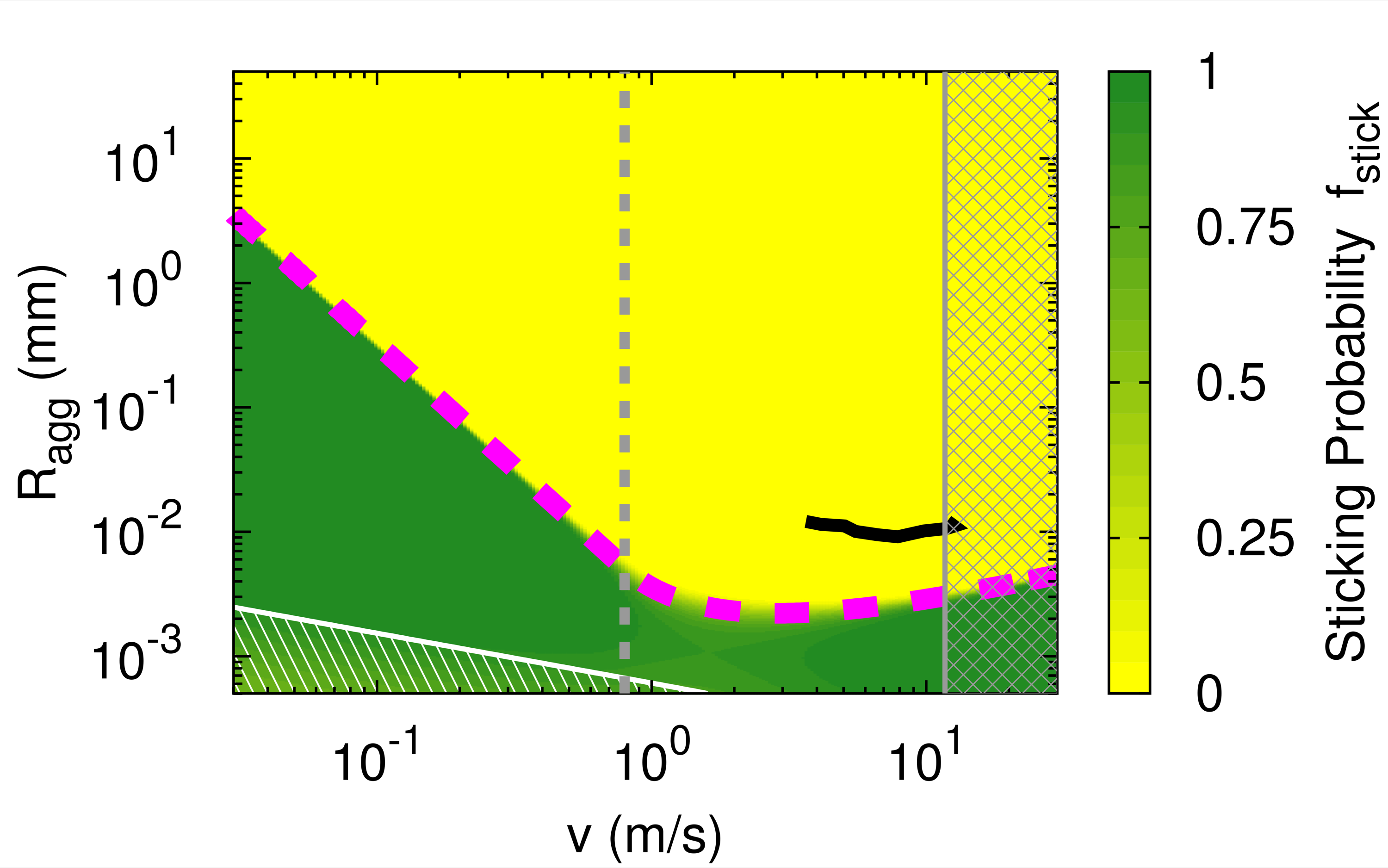}{\columnwidth}{(a) $\phi = 0.4$, $\alpha = 2$, and $\beta = 0.5$}
          \fig{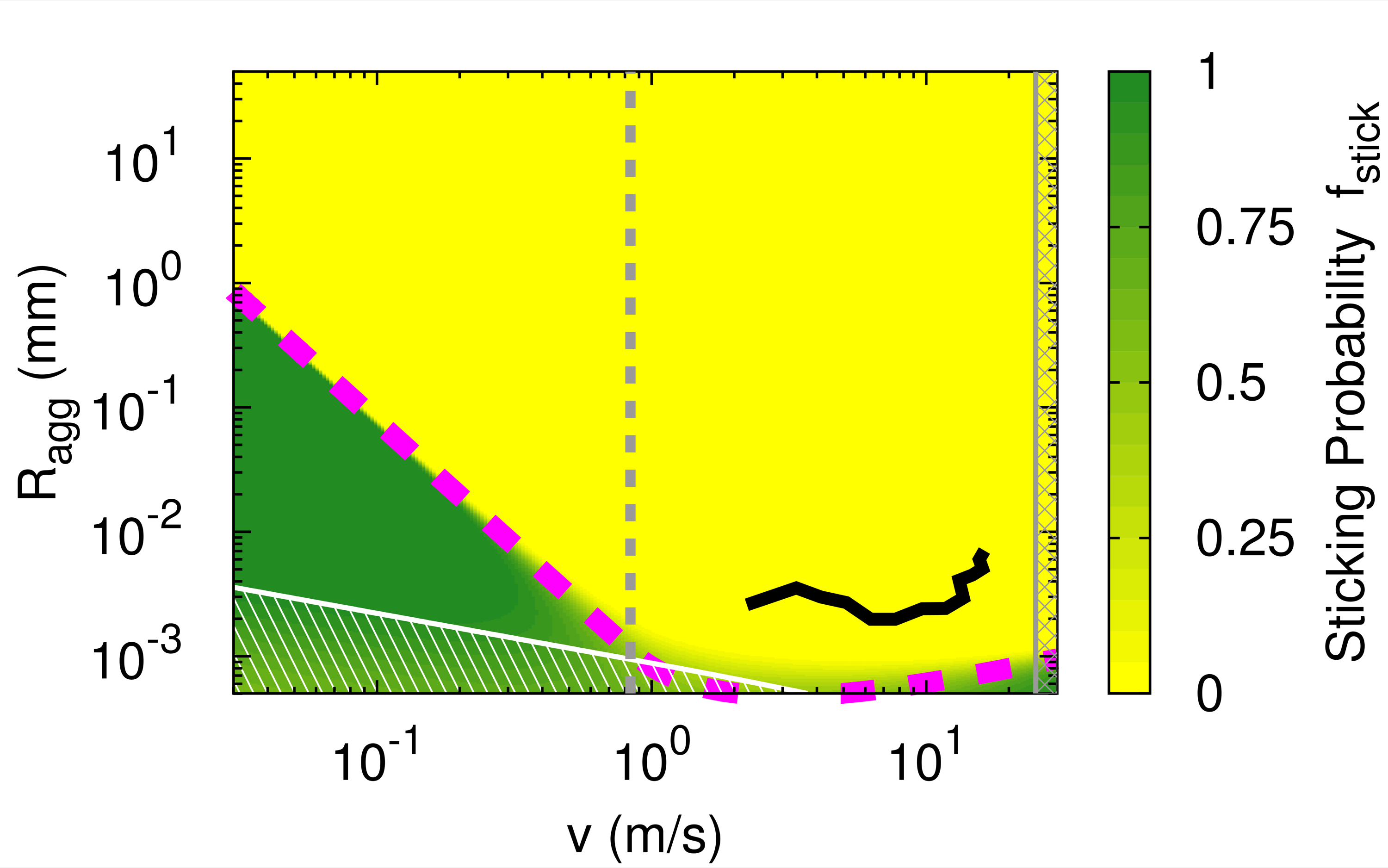}{\columnwidth}{(b) $\phi = 0.5$, $\alpha = 2$, and $\beta = 0.5$}}
\gridline{\fig{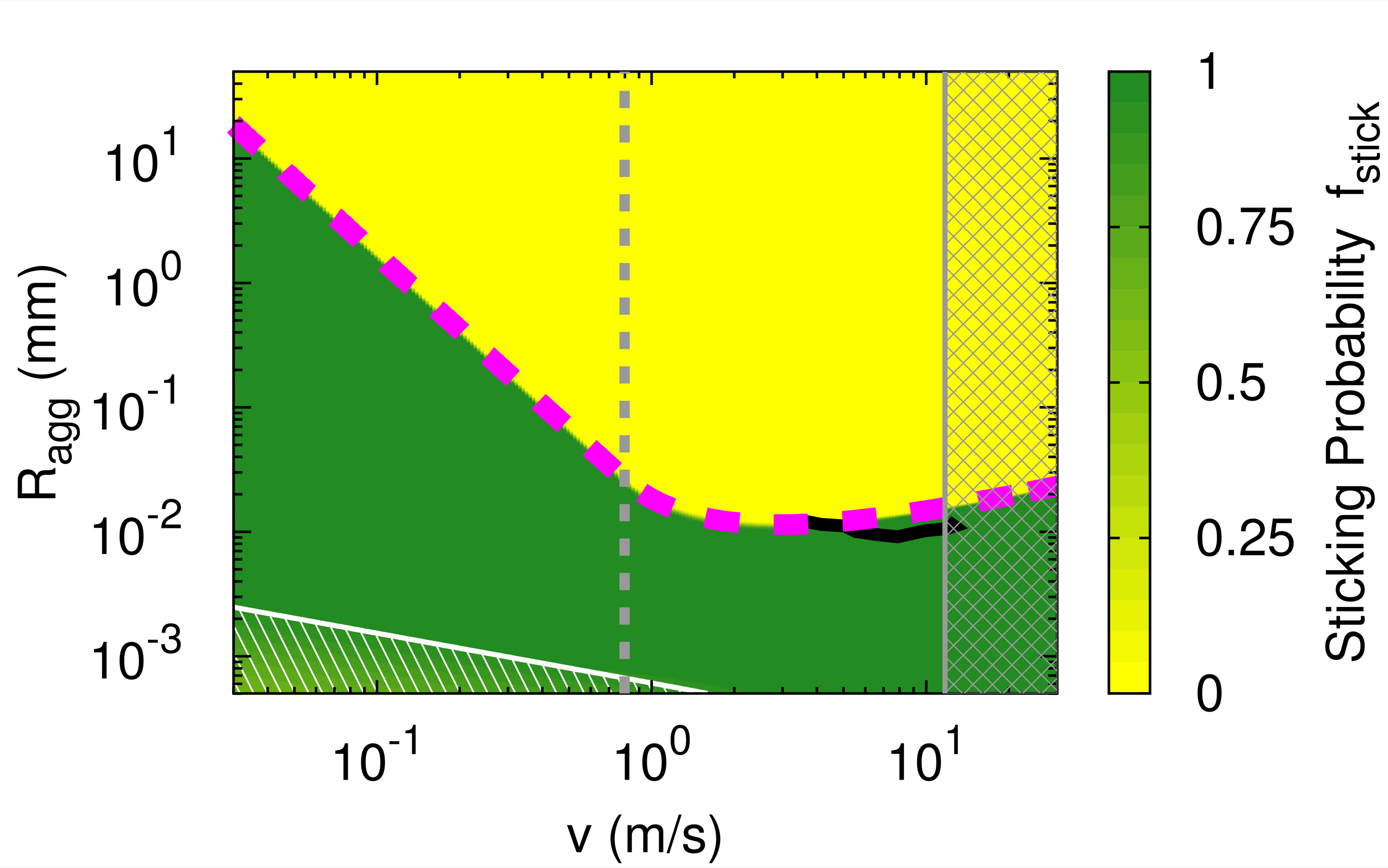}{\columnwidth}{(c) $\phi = 0.4$, $\alpha = 2$, and $\beta = 1.3$}
          \fig{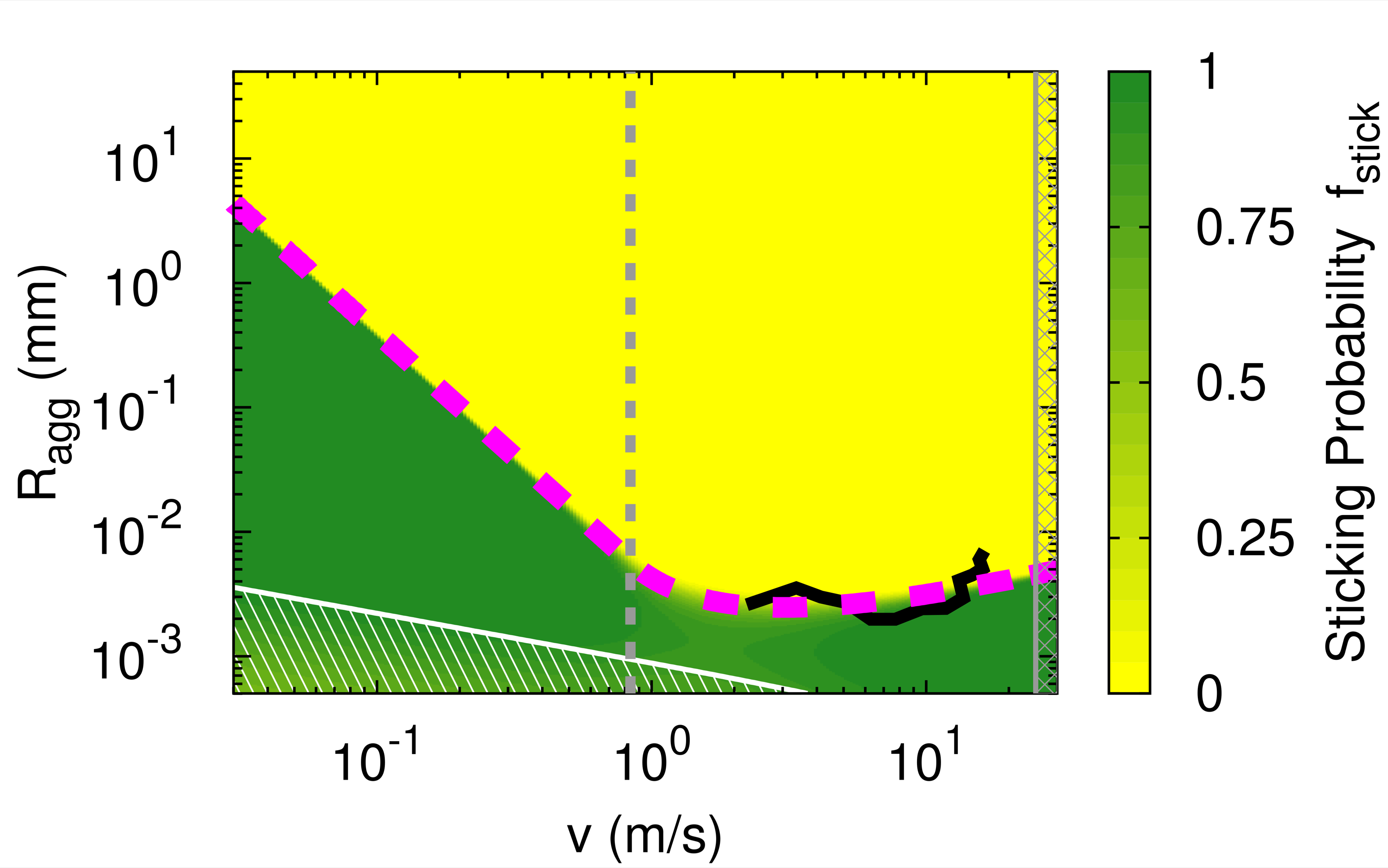}{\columnwidth}{(d) $\phi = 0.5$, $\alpha = 2$, and $\beta = 1.3$}}
\gridline{\fig{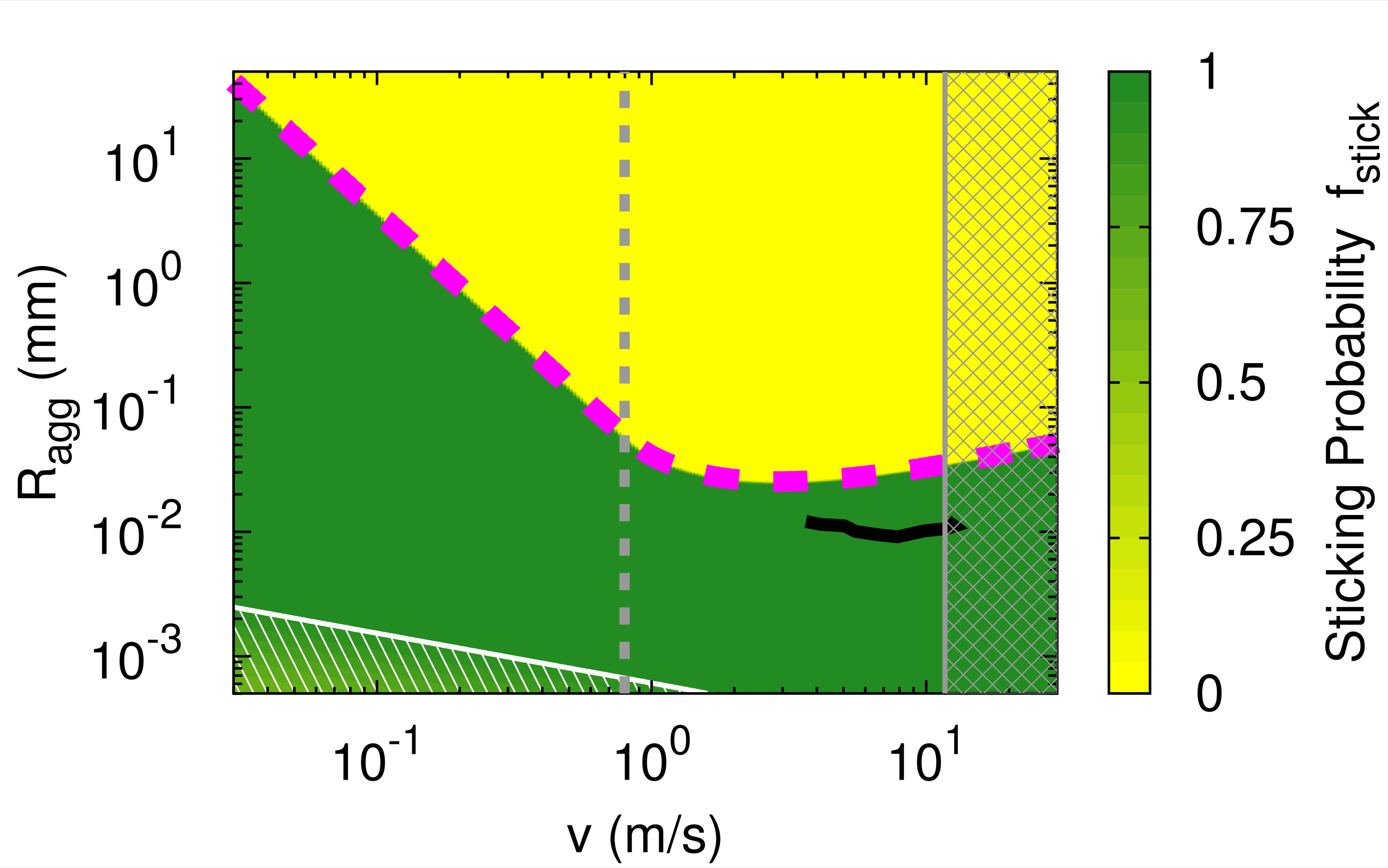}{\columnwidth}{(e) $\phi = 0.4$, $\alpha = 2$, and $\beta = 2$}
          \fig{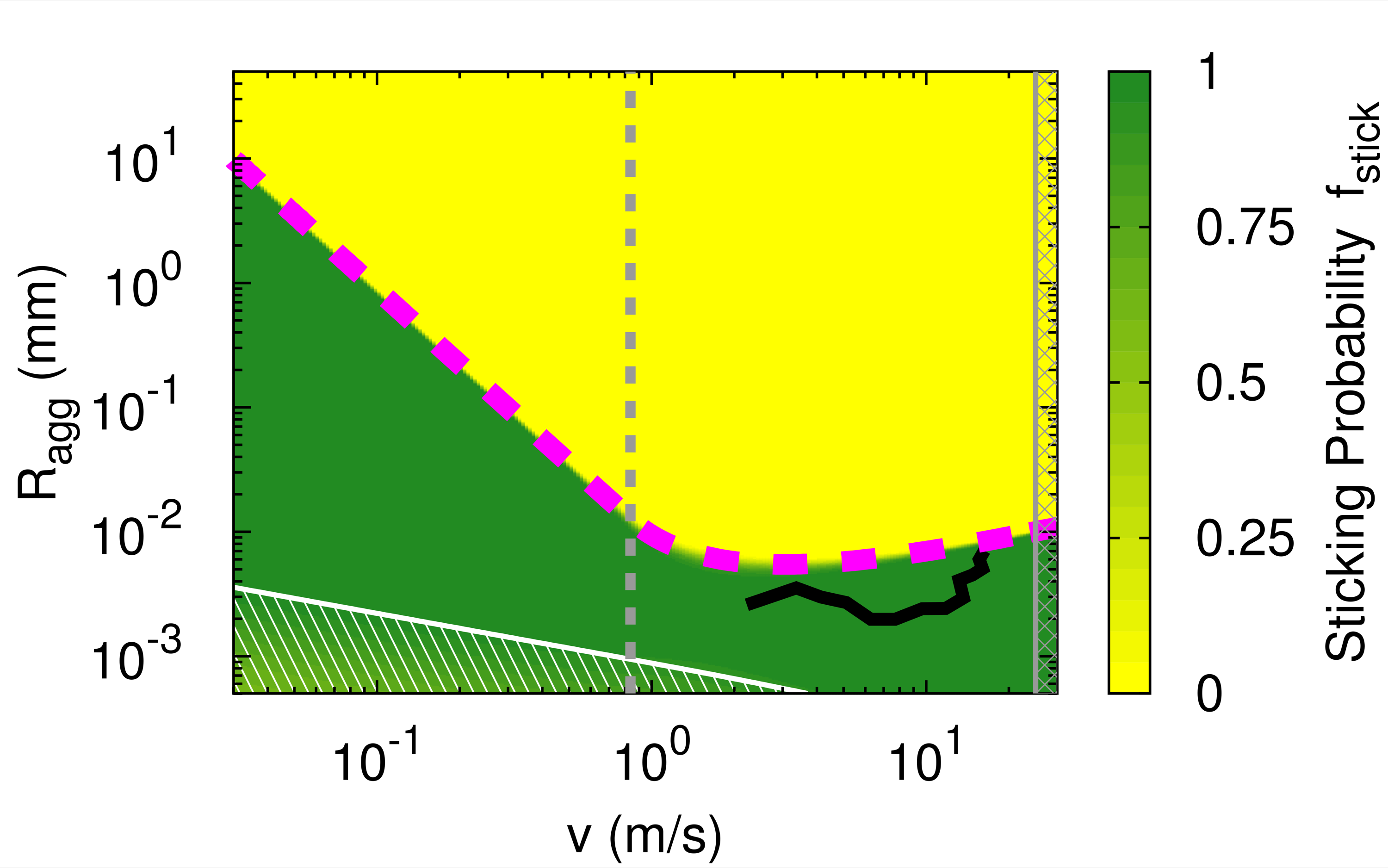}{\columnwidth}{(f) $\phi = 0.5$, $\alpha = 2$, and $\beta = 2$}}
\caption{
Sticking probability $f_{\rm stick}$ in the $v$--$R_{\rm agg}$ plane.
Here we fix $\alpha = 2$.
Each panel shows the result for a different set of $\phi$ and $\beta$.
The line styles and hatched regions are the same as in Figure \ref{fig:alpha}.
Among the parameter sets explored in this study, the combination of $\alpha = 2$ and $\beta = 1.3$ provides the most reasonable description of the simulation results of \citet{2025ApJ...983...75O}, as shown in panels (c) and (d).
}
\label{fig:beta}
\end{figure*}

The magenta dashed curves in Figure \ref{fig:alpha} show the condition $f_{\rm stick} = 50\%$ predicted by our model.
We denote the aggregate radius satisfying this condition at a given collision velocity by $R_{50}$.
The black lines show the $50\%$ sticking condition obtained from the numerical simulations of \citet{2025ApJ...983...75O}.
The vertical dashed lines indicate the critical velocity $v_{\rm crit}$ for the onset of plastic deformation.
The vertical solid lines indicate the velocity $v_{\rm A}$ at which $\delta_{\rm max} = R^{*}$.
For $v > v_{\rm A}$, $\delta_{\rm max}$ exceeds $R^{*}$, and the applicability of the elastoplastic contact model \citep{doi:10.1080/14786443008565033}, which assumes small deformation of colliding spheres, becomes uncertain.
We therefore indicate this region by cross-hatching in Figure \ref{fig:alpha}.
The white solid lines indicate the contours where $N_{\rm h} = 1$.
Our model for estimating ${\left\langle E_{\rm break} \right\rangle}$ is not applicable for $N_{\rm h} < 1$ (see Section \ref{sec:weakest}), and this region is also hatched in Figure \ref{fig:alpha}.

Figure \ref{fig:alpha} shows that the sticking--bouncing boundary depends strongly on $\alpha$.
As discussed in Section~3.1, for fixed collision velocity and filling factor, ${\left\langle E_{\rm break} \right\rangle} / K_{\rm rep} \propto {R_{\rm agg}}^{- 1 / \alpha}$.
Thus, $\alpha$ controls how strongly the relative fracture energy depends on aggregate size.
As $\alpha$ increases, the bond-breaking energies become more homogeneous and the $50\%$ sticking radius $R_{50}$ shifts to larger values.

Comparison with the numerical results of \citet{2025ApJ...983...75O} shows that, for $\beta = 1$, the cases with $\alpha = 1$ and $\alpha = 3$ do not reproduce the simulated sticking--bouncing boundary well.
For $\alpha = 1$, the predicted $R_{50}$ is systematically smaller than that inferred from the simulations, and the model underestimates $f_{\rm stick}$.
For $\alpha = 3$, the predicted $R_{50}$ is systematically larger than that inferred from the simulations, and the model overestimates $f_{\rm stick}$.
In contrast, for $\alpha = 2$, the $f_{\rm stick} = 50\%$ boundary shown by the magenta dashed curve is broadly consistent with the black line obtained from the simulations of \citet{2025ApJ...983...75O}.
Thus, for $\beta = 1$, $\alpha \approx 2$ provides the best overall agreement with the numerical results.

Figure \ref{fig:alpha} also shows that, in the elastic regime of $v \le v_{\rm crit}$, the $50\%$ sticking radius $R_{50}$ can be described by a power-law function of collision velocity $v$.
It is also clear that the power-law exponent depends on $\alpha$.
Here, we derive this exponent from the scaling relations obtained in Section \ref{sec:dependence}.
In the elastic collision regime, Equations \eqref{eq:ratio_R_agg} and \eqref{eq:ratio_v_1} give
\begin{equation}
\frac{ {\left\langle E_{\rm break} \right\rangle} }{K_{\rm rep}} \propto {R_{\rm agg}}^{- 1 / \alpha} \times v^{- 4/5 - 2 / {\left( 5 \alpha \right)}}.
\label{eq:r_agg_v}
\end{equation}
The sticking probability $f_{\rm stick}$ is given by Equation \eqref{eq:f_stick}, and the condition $f_{\rm stick} = 50\%$ is equivalent to ${\left\langle E_{\rm break} \right\rangle} / K_{\rm rep} = 1$.
Substituting this condition into the scaling relation \eqref{eq:r_agg_v} yields
\begin{equation}
R_{50} \propto v^{- 2/5 - 4 \alpha / 5}.
\end{equation}
For example, when $\alpha = 2$, this scaling gives $R_{50} \propto v^{-2}$.
A larger $\alpha$ gives a steeper dependence of $R_{50}$ on $v$.

\subsection{Dependence on $\beta$}
\label{sec:beta}

In Section \ref{sec:beta}, we examine how the parameter $\beta$, which controls the magnitude of the bond-breaking energy, affects the sticking probability.
Here, we fix $\alpha = 2$, which provided the best agreement with the numerical results of \citet{2025ApJ...983...75O} (see Section \ref{sec:alpha}), and vary $\beta$.
The $\beta$ dependence for $\alpha = 1$ and $\alpha = 3$ is discussed in Appendix \ref{app:1_and_3}.

Figure \ref{fig:beta} shows the sticking probability $f_{\rm stick}$ for different values of $\beta$ for $\phi = 0.4$ and $\phi = 0.5$.
We find that $R_{50}$ increases with increasing $\beta$.
This is because increasing $\beta$ increases the characteristic breaking energy of a single bond, $E_{1}$, through Equation \eqref{eq:E_1}, thereby increasing the fracture energy of the entire contact region.
A larger $\beta$ means that more energy is required to break the contact region, and thus bouncing occurs only for larger aggregates at a given collision velocity.
As a result, the $f_{\rm stick} = 0.5$ boundary shifts toward larger $R_{\rm agg}$.

Comparison with the numerical results of \citet{2025ApJ...983...75O} shows that, for $\alpha = 2$, the cases with $\beta = 0.5$ and $\beta = 2$ do not reproduce the simulated sticking--bouncing boundary for either $\phi = 0.4$ or $\phi = 0.5$.
For $\beta = 0.5$, the breaking energy of a single bond is underestimated, and the contact region becomes too easy to break.
In contrast, for $\beta = 2$, the fracture energy of the contact region becomes too large.
For $\beta = 1.3$, however, the $f_{\rm stick} = 0.5$ boundary predicted by the model is broadly consistent with the numerical results of \citet{2025ApJ...983...75O} for both $\phi = 0.4$ and $\phi = 0.5$.
Thus, for $\alpha = 2$, $\beta \simeq 1.3$ best reproduces the sticking--bouncing boundary obtained in the numerical simulations.

In Appendix \ref{app:1_and_3}, we also show the $\beta$ dependence for $\alpha = 1$ and $\alpha = 3$.
For these values of $\alpha$, no single value of $\beta$ can reproduce the numerical results of \citet{2025ApJ...983...75O} for both $\phi = 0.4$ and $\phi = 0.5$.
This suggests that, among the parameter sets explored here, the combination of $\alpha = 2$ and $\beta = 1.3$ provides the most reasonable description of the simulation results.

\section{Discussion}
\label{sec:discussion}

In this study, we constructed a semi-analytic model for the sticking probability of low-velocity collisions between dust aggregates.
In the model, the size of the contact region formed at maximum compression and the repulsive energy available after maximum compression are calculated from the elastoplastic contact model (see Section \ref{sec:macro}).
The fracture energy required to separate the contact region is then evaluated using weakest-link statistics (see Section \ref{sec:micro}).
In Section \ref{sec:results}, we showed that the parameter set with $\alpha = 2$ and $\beta = 1.3$ provides a reasonable description of the simulation results of \citet{2025ApJ...983...75O} among the cases explored here.
In this section, we apply this best-fit parameter set to dust growth in protoplanetary disks.

Astronomical observations with the Atacama Large Millimeter/submillimeter Array (ALMA) and the Very Large Array (VLA), especially millimeter-wave polarization observations and multi-frequency analyses, have provided constraints on the sizes and internal structures of dust aggregates in protoplanetary disks \citep[e.g.,][]{2019ApJ...885...52T, 2024ARA&A..62..157B, 2025ApJ...980...50Y, 2025A&A...702A..56Z}.
In particular, radiative-transfer modeling of disk polarization can constrain the 
filling factors of dust aggregates \citep[e.g.,][]{2019ApJ...885...52T}.

Recently, \citet{2024NatAs...8.1148U} suggested that moderately porous aggregates with $R_{\rm agg} \approx 0.1$--$1~{\rm mm}$ and $\phi \approx 0.2$--$0.3$ can explain both the thermal emission and polarization observed in the protoplanetary disk around the T Tauri star IM Lup.
They also inferred that the collision velocities of these aggregates are $v \lesssim 1~{\rm m}~{\rm s}^{-1}$.
These observationally inferred aggregate sizes, filling factors, and collision velocities provide important constraints for testing whether dust growth is limited by the bouncing barrier.

\begin{figure*}[!t]
\centering
\gridline{\fig{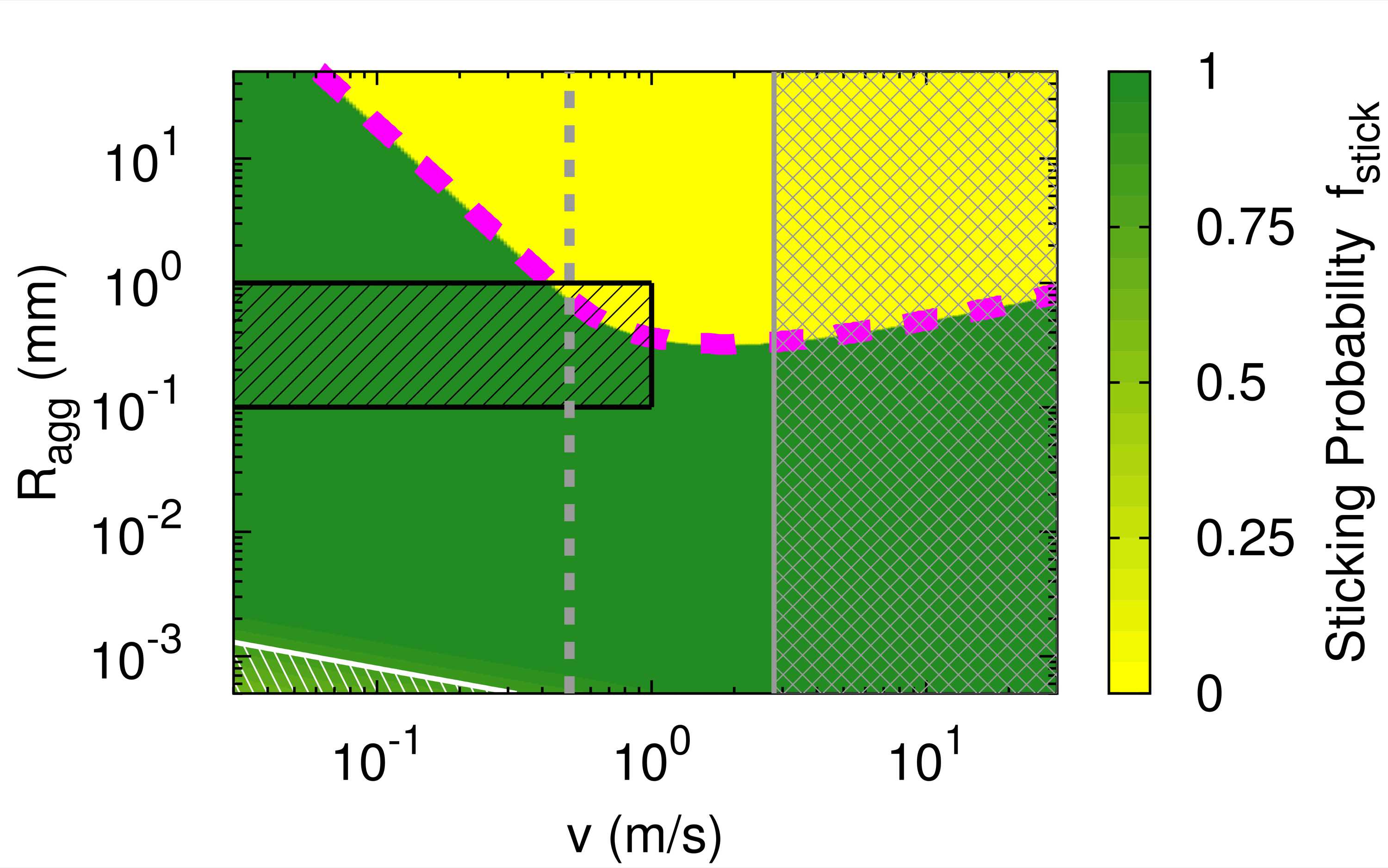}{\columnwidth}{(a) $\phi = 0.2$, $\alpha = 2$, and $\beta = 1.3$}
          \fig{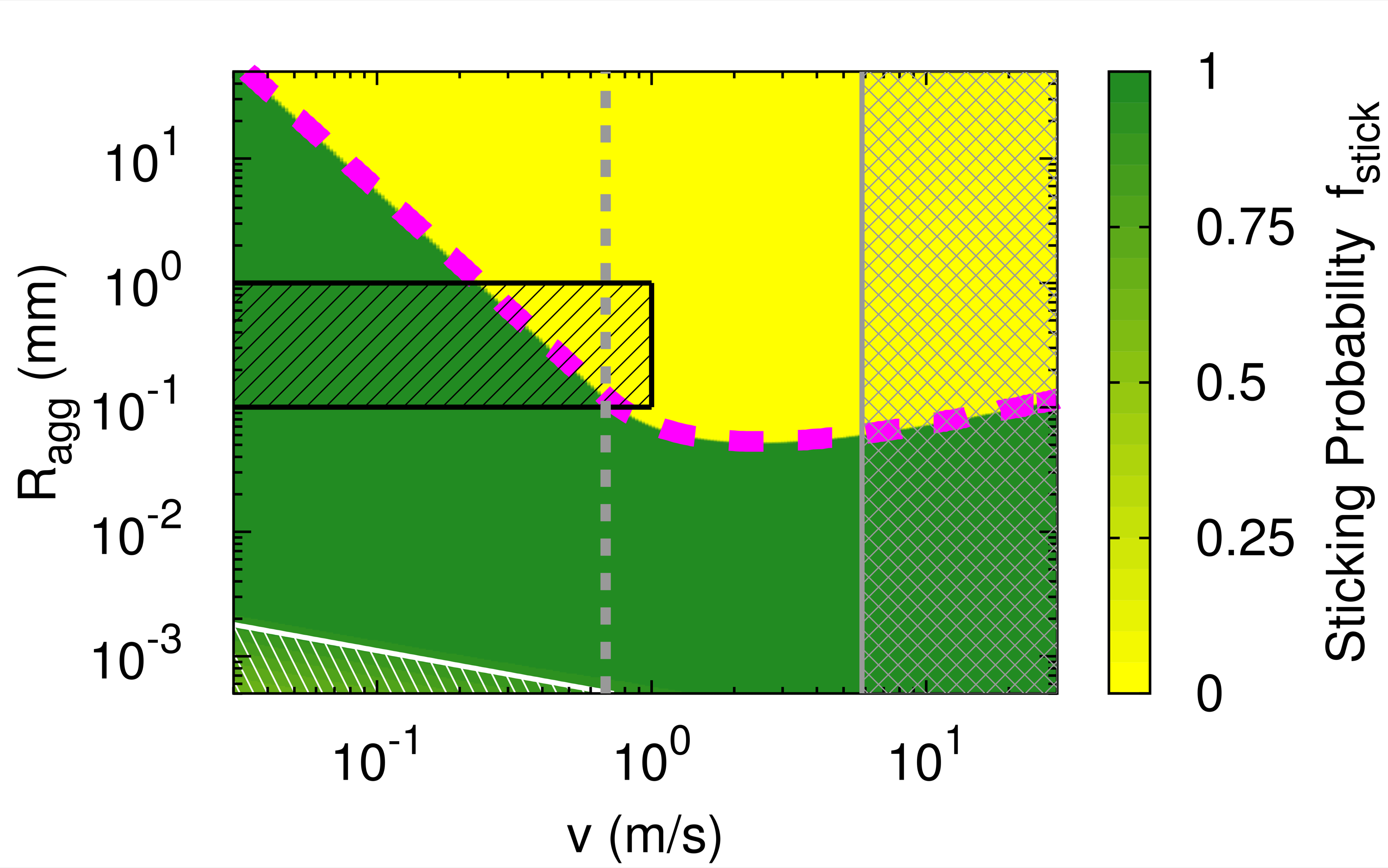}{\columnwidth}{(b) $\phi = 0.3$, $\alpha = 2$, and $\beta = 1.3$}}
\caption{
Sticking probability $f_{\rm stick}$ in the $v$--$R_{\rm agg}$ plane for (a) $\phi = 0.2$ and (b) $\phi = 0.3$.
We adopt $\alpha = 2$ and $\beta = 1.3$, corresponding to the best-fit parameter set obtained in Section \ref{sec:results}.
The black hatched regions indicate the parameter space inferred from observations of the IM Lup disk \citep{2024NatAs...8.1148U}.
The other line styles and hatched regions are the same as in Figure \ref{fig:alpha}.
}
\label{fig:porous}
\end{figure*}

Figure \ref{fig:porous} shows the sticking probability for $\phi = 0.2$ and $\phi = 0.3$, calculated using the best-fit parameter set obtained in Section \ref{sec:results}, namely $\alpha = 2$ and $\beta = 1.3$.
The black hatched regions indicate the parameter space inferred from observations of the IM Lup disk \citep{2024NatAs...8.1148U}: $0.1~{\rm mm} \le R_{\rm agg} \le 1~{\rm mm}$ and $v \lesssim 1~{\rm m}~{\rm s}^{-1}$.
For both filling factors, there exist collision velocities with $v < 1~{\rm m}~{\rm s}^{-1}$ at which the $50\%$ sticking radius lies in the range $0.1~{\rm mm} \le R_{50} \le 1~{\rm mm}$.
This result suggests that the observed (sub)millimeter-sized dust aggregates may have reached sizes limited by the sticking--bouncing boundary.

Several caveats should be noted.
First, the parameters $\alpha$ and $\beta$ were constrained through comparison with DEM simulations of relatively compact aggregates with $\phi = 0.4$ and $0.5$.
Whether the same parameter values can be directly applied to more porous aggregates with $\phi = 0.2$--$0.3$ should be tested in future DEM simulations.
Second, the present model assumes equal-mass, head-on collisions between non-sintered icy aggregates.
In actual protoplanetary disks, however, the sticking--bouncing boundary may be affected by the impact parameter, mass ratio, aggregate shape and structure, composition of the constituent grains, and sintering \citep[e.g.,][]{2008ApJ...675..764L, 2013Icar..225...75K, 2017ApJ...841...36S}.
Third, our model focuses on the sticking--bouncing outcome of aggregate collisions and does not treat gradual mass loss through erosion, i.e., the production of small fragments during collisions.
Because dust aggregates in protoplanetary disks undergo a large number of collisions, erosive collisions may influence the resulting size and filling-factor distributions \citep[e.g.,][]{2015A&A...574A..83K}.
These effects should therefore be investigated in future work.

Despite these limitations, our model predicts a sticking--bouncing boundary that is broadly consistent with the observationally inferred aggregate sizes, filling factors, and collision velocities \citep{2024NatAs...8.1148U}.
This agreement suggests that the bouncing barrier may play an important role in setting the dust-size distribution in protoplanetary disks.
Because the present model is semi-analytic, it can be applied to millimeter-sized or larger aggregates that are difficult to simulate directly with DEM simulations.
It may therefore provide a useful framework for future dust growth calculations and for interpreting observations of protoplanetary disks.

\section{Summary}

In this study, we develop a semi-analytic model for the sticking--bouncing boundary of dust aggregates.
In the model, the compression phase of an aggregate collision is described using an elastoplastic contact model \citep{doi:10.1080/14786443008565033, 2024arXiv240815573A}, which gives the maximum contact radius and the repulsive energy available after compression (Section \ref{sec:macro}).
The subsequent separation phase is modeled as fracture of an interparticle bond network in the contact region (Section \ref{sec:micro}), and the fracture energy is evaluated using weakest-link statistics \citep{Weibull1939}.
The sticking probability $f_{\rm stick}$ is then determined by comparing the fracture energy $E_{\rm break}$ required to break the contact region with the repulsive energy $K_{\rm rep}$ available for 
separation (Section \ref{sec:f_stick}).

This framework naturally explains why larger aggregates are more likely to bounce: larger contact regions are statistically more likely to contain weak bonds.
Comparison with DEM simulations by \citet{2025ApJ...983...75O} shows that the model reproduces the simulated sticking--bouncing boundary reasonably well for $\alpha \approx 2$ and $\beta \approx 1.3$ (Section \ref{sec:beta}).
When this best-fit parameter set is applied to moderately porous aggregates inferred from ALMA observations of the IM Lup disk \citep{2024NatAs...8.1148U}, the predicted sticking--bouncing boundary is broadly consistent with the observationally inferred aggregate sizes, filling factors, and collision velocities (Section \ref{sec:discussion}).
This result suggests that our semi-analytic model may be useful not only for interpreting collisional outcomes in numerical simulations, but also for predicting the collisional evolution of protoplanetary dust aggregates.

In future studies, we plan to extend this approach to dust aggregates with compositions other than ice by performing DEM simulations and estimating the model parameters $\alpha$ and $\beta$ within the framework of the present semi-analytic model.
A particularly important target is silicate dust aggregates, because their sticking--bouncing boundary is directly relevant to planetesimal formation in the terrestrial planet-forming region of the Solar System and, ultimately, to the origin of the Earth.
In addition, a large number of laboratory experiments have been performed for silicate dust aggregates \citep[e.g.,][]{2008ApJ...675..764L, 2010A&A...513A..56G, 2013Icar..225...75K}, and these data may allow us to constrain the model parameters $\alpha$ and $\beta$ independently of DEM simulations.
This will be examined in a forthcoming study.

\begin{acknowledgments}
This work was supported by JSPS KAKENHI Grant Numbers JP24K17118, JP24KK0072, JP25K00025, JP25K01063, JP25K07392, JP25H00678, JP26KJ1168, and JP26K06964.
We acknowledge the use of ChatGPT for assistance with preparing schematic illustrations, checking English grammar, and improving the clarity of the text.
\end{acknowledgments}

\appendix

\section{Yield Stress and Reduced Young's Modulus of Aggregates}
\label{app:yield}

\begin{figure*}[]
\centering
\gridline{\fig{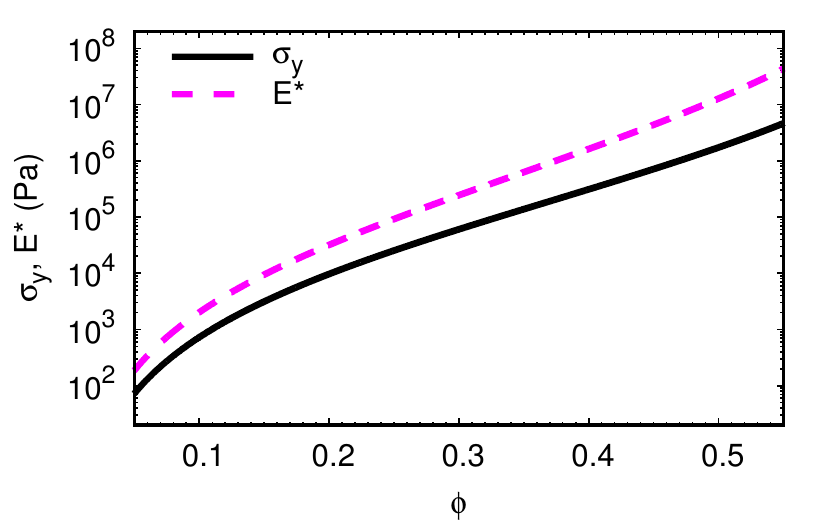}{\columnwidth}{(a) $\sigma_{\rm y}$ and $E^{*}$}
          \fig{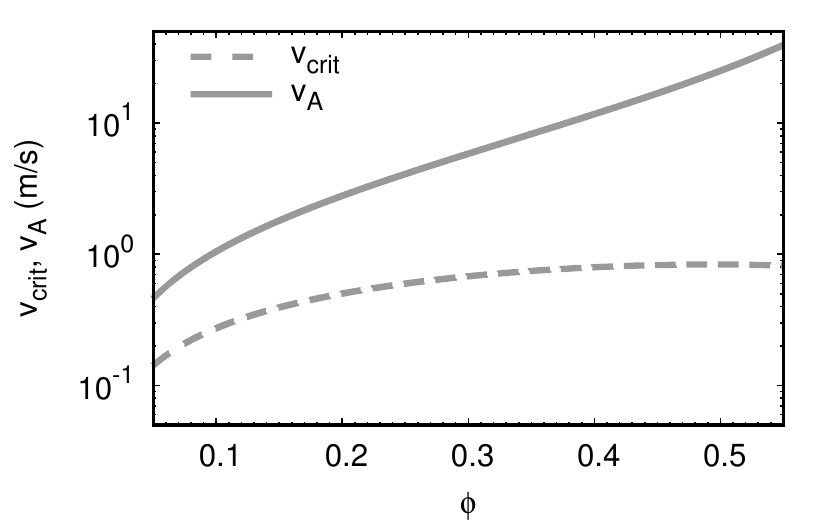}{\columnwidth}{(b) $v_{\rm crit}$ and $v_{\rm A}$}}
\caption{
(a) Filling-factor dependence of $\sigma_{\rm y}$ and $E^{*}$.
(b) Filling-factor dependence of $v_{\rm crit}$ and $v_{\rm A}$.
}
\label{fig:sigma_v}
\end{figure*}

We regard the aggregates as effective elastoplastic spheres with reduced Young's modulus $E^{*}$ and yield stress $\sigma_{\rm y}$.
\citet{2023ApJ...953....6T} reported the filling-factor dependence of the yield stress of icy dust aggregates composed of monodisperse constituent grains with radius $r_{1} = 100~{\rm nm}$.
We adopt their empirical formula,
\begin{equation}
\sigma_{\rm y} = C_{\rm y} {\left( \frac{1}{\phi} - \frac{1}{\phi_{\rm max}} \right)}^{-3},
\end{equation}
where $C_{\rm y} = 4.7 \times 10^{5}~{\rm Pa}$ and $\phi_{\rm max} = 0.74$.
When the pressure exerted on an aggregate is lower than $\sigma_{\rm y}$, the aggregate is assumed to behave elastically.
The bulk modulus $K$ in this elastic regime is then given by \citep[e.g.,][]{landau1986theory}
\begin{equation}
K = \phi \frac{\partial \sigma_{\rm y}}{\partial \phi},
\end{equation}
where we identify the isotropic pressure with the yield stress $\sigma_{\rm y}$.
Using the Poisson's ratio of aggregates, $\nu$, the reduced Young's modulus of the aggregates is written as \citep[e.g.,][]{landau1986theory}
\begin{equation}
E^{*} = \frac{3 {\left( 1 - 2 \nu \right)}}{2 {\left( 1 - \nu^{2} \right)}} K.
\end{equation}
In this study, we adopt $\nu = 0.25$ as a fiducial value.
Figure \ref{fig:sigma_v}(a) shows the filling-factor dependence of the yield stress $\sigma_{\rm y}$ and the reduced Young's modulus $E^{*}$.

Furthermore, substituting $\sigma_{\rm y}$ and $E^{*}$ into Equation \eqref{eq:v_crit} yields the filling-factor dependence of the critical collision velocity for yielding, $v_{\rm crit}$.
Figure \ref{fig:sigma_v}(b) shows the resulting filling-factor dependence of $v_{\rm crit}$.

The elastoplastic contact model of \citet{doi:10.1080/14786443008565033} assumes small deformation of the colliding spheres.
Thus, its applicability is uncertain when $\delta_{\rm max} \gtrsim R^{*}$.
By substituting $\delta_{\rm max} = R^{*}$ into Equation \eqref{eq:delta_max_2}, we define the collision velocity $v_{\rm A}$ at which $\delta_{\rm max} = R^{*}$.
This velocity provides an approximate upper limit for the applicability of the elastoplastic contact model.
Using the filling-factor dependence of $\sigma_{\rm y}$ and $E^{*}$, $v_{\rm A}$ can also be expressed as a function of $\phi$; this dependence is shown in Figure \ref{fig:sigma_v}(b).

\section{Mean Coordination Number}
\label{app:Z}

In Section \ref{sec:micro}, we constructed a semi-analytic model for the total fracture energy $E_{\rm break}$.
In this model, the aggregate--aggregate contact region is approximated as a bond network.
We assumed that $N_{\rm S}$ chains are connected in parallel and that $N_{\rm S}$ is proportional to the mean coordination number of the aggregates, $Z$ (Equation \eqref{eq:N_S}).

\citet{2019Icar..324....8A} showed that the mean coordination number $Z$ can be expressed as a function of the filling factor $\phi$ as
\begin{equation}
Z = 2 + 9.38 \phi^{1.62}.
\end{equation}
We adopt this relation in the present study.

\section{Material Properties of Constituent Grains}
\label{app:material}

The DEM simulations of \citet{2025ApJ...983...75O} assumed monodisperse icy constituent grains with radius $r_{1} = 100~{\rm nm}$ and material density $\rho_{\rm mat} = 1000~{\rm kg}~{\rm m}^{-3}$.
This setup is common to the other DEM simulations referred to in this study \citep{2023ApJ...951L..16A, 2024arXiv240815573A, 2023ApJ...953....6T, 2025ApJ...995..207T}, and the relevant material properties are summarized by \citet{2007ApJ...661..320W}.

\citet{2007ApJ...661..320W} modeled submicron-sized ice grains as adhesive elastic spheres.
The normal interaction between grains is described by the Johnson--Kendall--Roberts (JKR) contact model \citep{1971RSPSA.324..301J}.
For two icy grains with radius $r_{1} = 100~{\rm nm}$, the energy required to break the contact is $E_{\rm b} = 6.1 \times 10^{-17}~{\rm J}$.

In these DEM simulations, tangential interactions between grains, including rolling, sliding, and twisting, are also taken into account.
For these tangential motions, \citet{2007ApJ...661..320W} constructed a model in which the interaction behaves as a linear spring for small displacements, whereas frictional energy dissipation occurs for large displacements.
According to \citet{2025ApJ...983...75O}, energy dissipation during the separation phase is mainly caused by rolling motion between grains.
In their DEM simulations, the energy dissipated when one grain rolls by $90^{\circ}$ over another, $E_{\rm r}$, is assumed to be $E_{\rm r} = 4.7 \times 10^{-16}~{\rm J}$.

During the separation phase, the configuration of grains changes before interparticle bonds are broken, and this rearrangement is accompanied by energy dissipation through rolling motion \citep{2025ApJ...983...75O}.
The parameter $\beta$ in Equation \eqref{eq:E_1} is therefore an order-unity factor that represents the typical amount of rolling displacement before bond breaking.

\section{Dependence of the Sticking Probability on $\beta$ for $\alpha = 1$ and $\alpha = 3$}
\label{app:1_and_3}

\begin{figure*}[]
\centering
\gridline{\fig{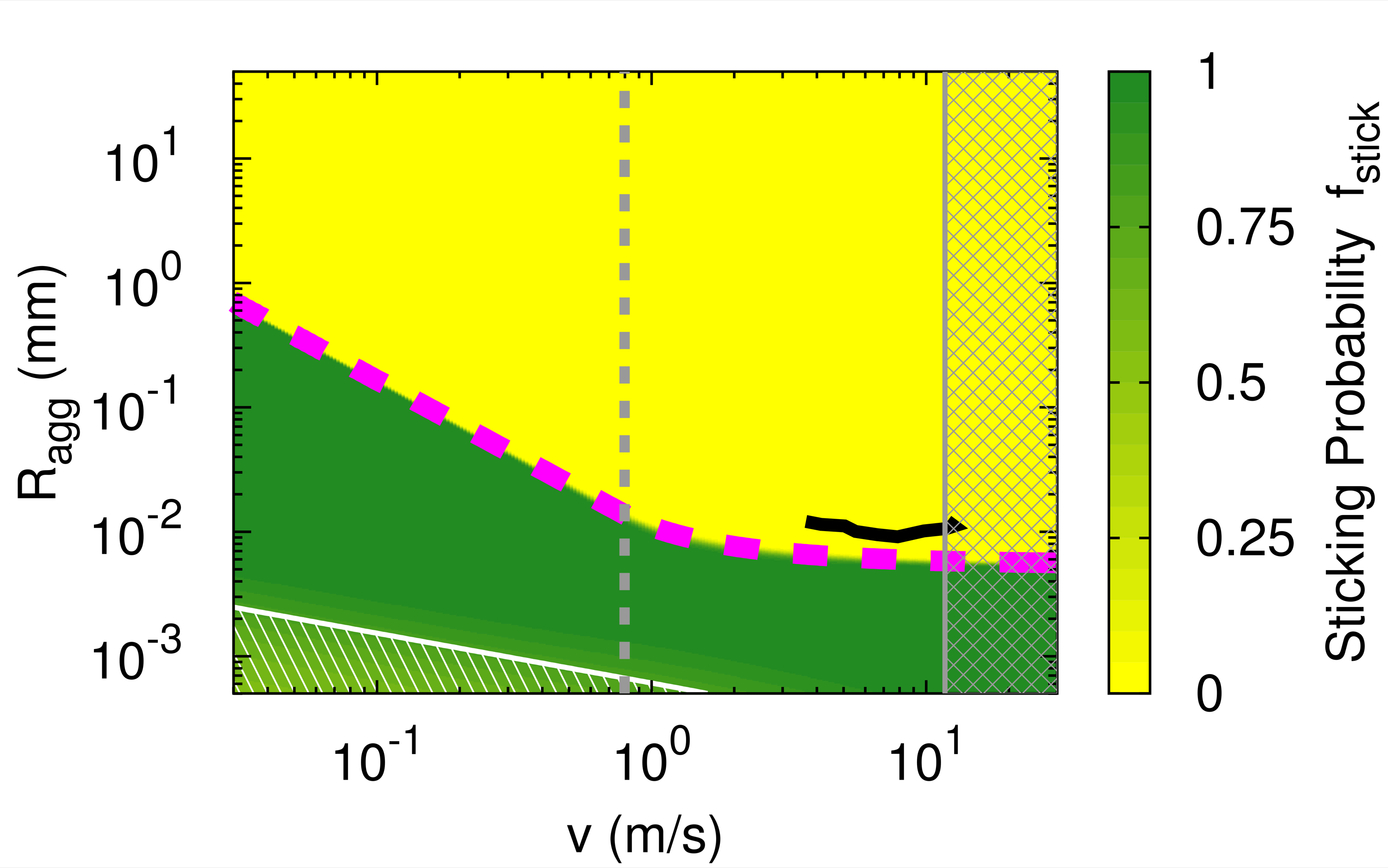}{\columnwidth}{(a) $\phi = 0.4$, $\alpha = 1$, and $\beta = 4$}
          \fig{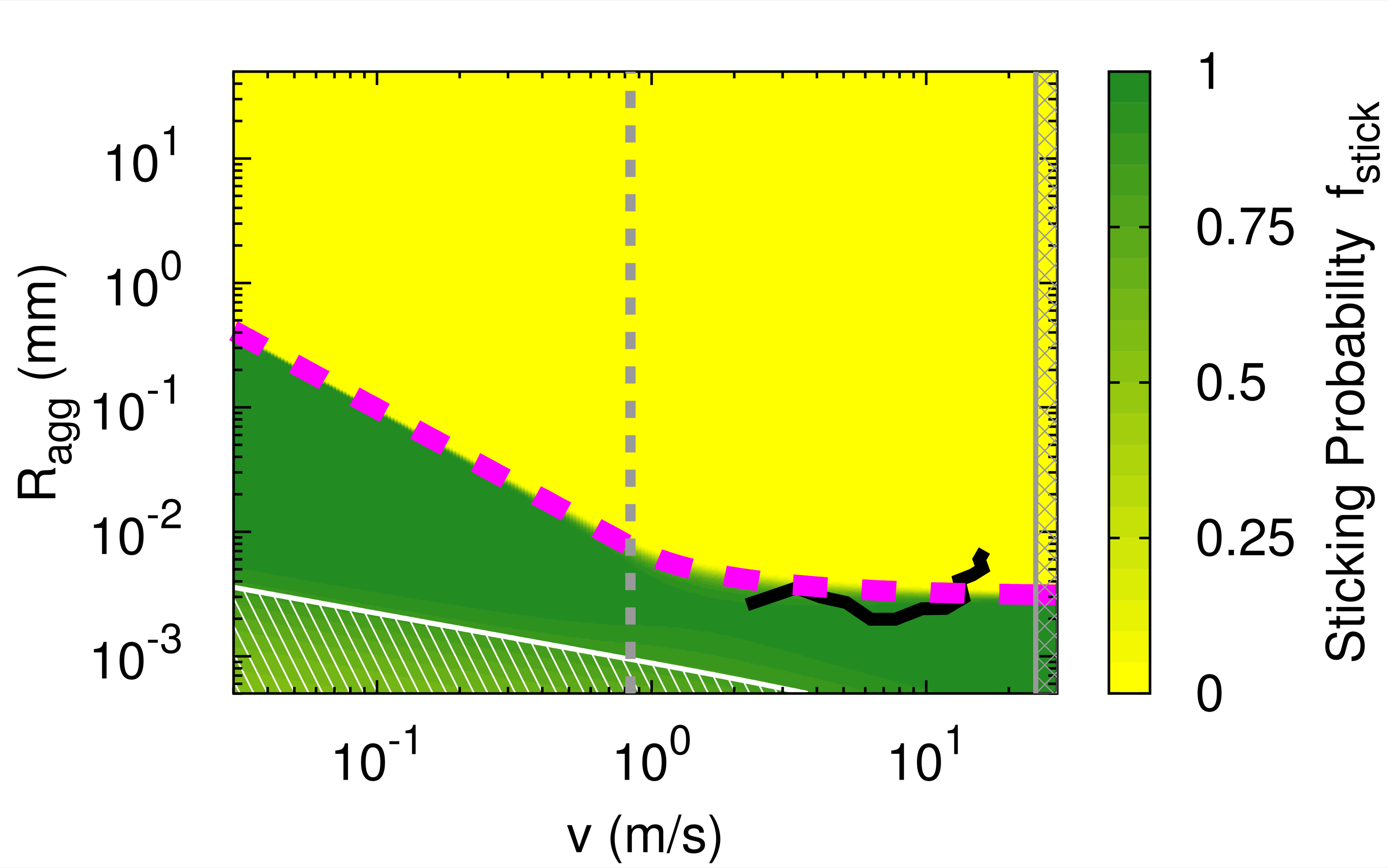}{\columnwidth}{(b) $\phi = 0.5$, $\alpha = 1$, and $\beta = 4$}}
\gridline{\fig{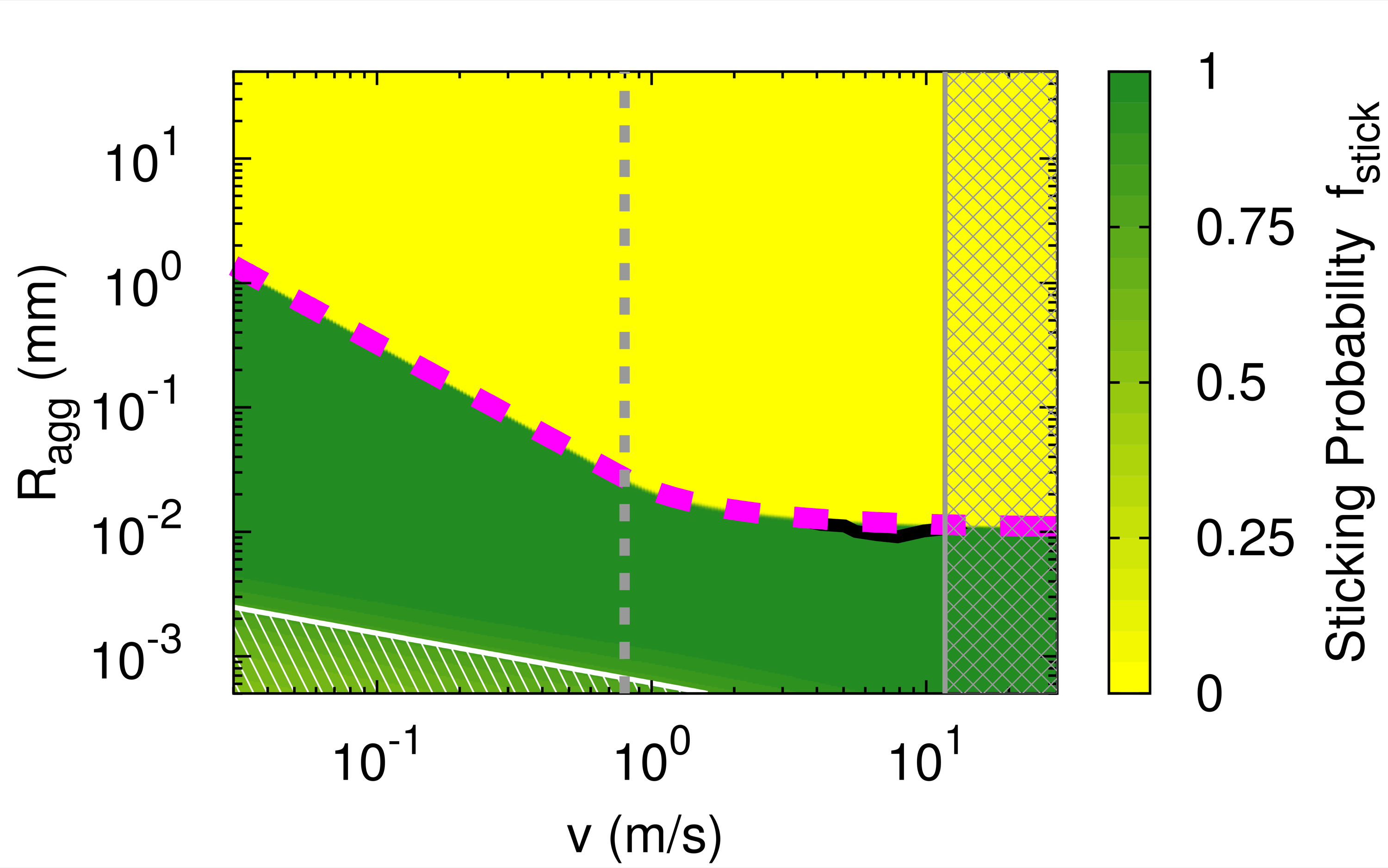}{\columnwidth}{(c) $\phi = 0.4$, $\alpha = 1$, and $\beta = 8$}
          \fig{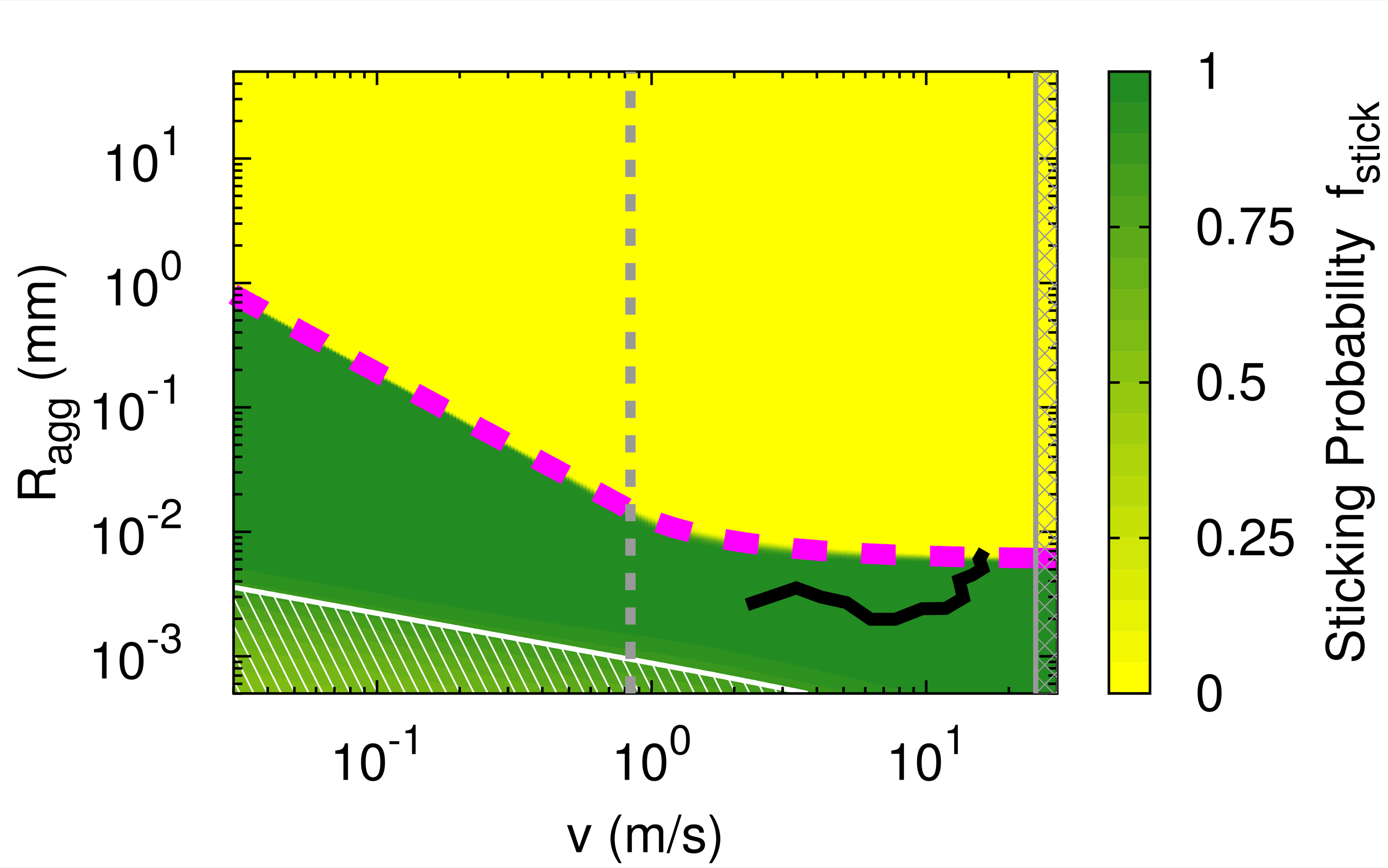}{\columnwidth}{(d) $\phi = 0.5$, $\alpha = 1$, and $\beta = 8$}}
\caption{
Sticking probability $f_{\rm stick}$ in the $v$--$R_{\rm agg}$ plane.
Here we fix $\alpha = 1$.
Each panel shows the result for a different set of $\phi$ and $\beta$.
The line styles and hatched regions are the same as in Figure \ref{fig:alpha}.
}
\label{fig:alpha_1}
\end{figure*}

Figure \ref{fig:alpha_1} shows the $\beta$ dependence of $f_{\rm stick}$ for $\alpha = 1$ at $\phi = 0.4$ and $0.5$.
When $\alpha = 1$ and $\beta = 4$, the $f_{\rm stick} = 50\%$ sticking--bouncing boundary, $R_{50}$, predicted by our model agrees well with the numerical results of \citet{2025ApJ...983...75O} for $\phi = 0.5$ (Figure \ref{fig:alpha_1}(b)).
However, for $\phi = 0.4$, the predicted $R_{50}$ is smaller than the boundary obtained from their numerical simulations (Figure \ref{fig:alpha_1}(a)).
In contrast, when $\beta = 8$, the predicted $R_{50}$ agrees well with the numerical results for $\phi = 0.4$ (Figure \ref{fig:alpha_1}(c)), whereas it becomes larger than the boundary obtained from the simulations for $\phi = 0.5$ (Figure \ref{fig:alpha_1}(d)).
Thus, for $\alpha = 1$, no single value of $\beta$ can simultaneously reproduce the numerical results of \citet{2025ApJ...983...75O} for both $\phi = 0.4$ and $0.5$.

\begin{figure*}[]
\centering
\gridline{\fig{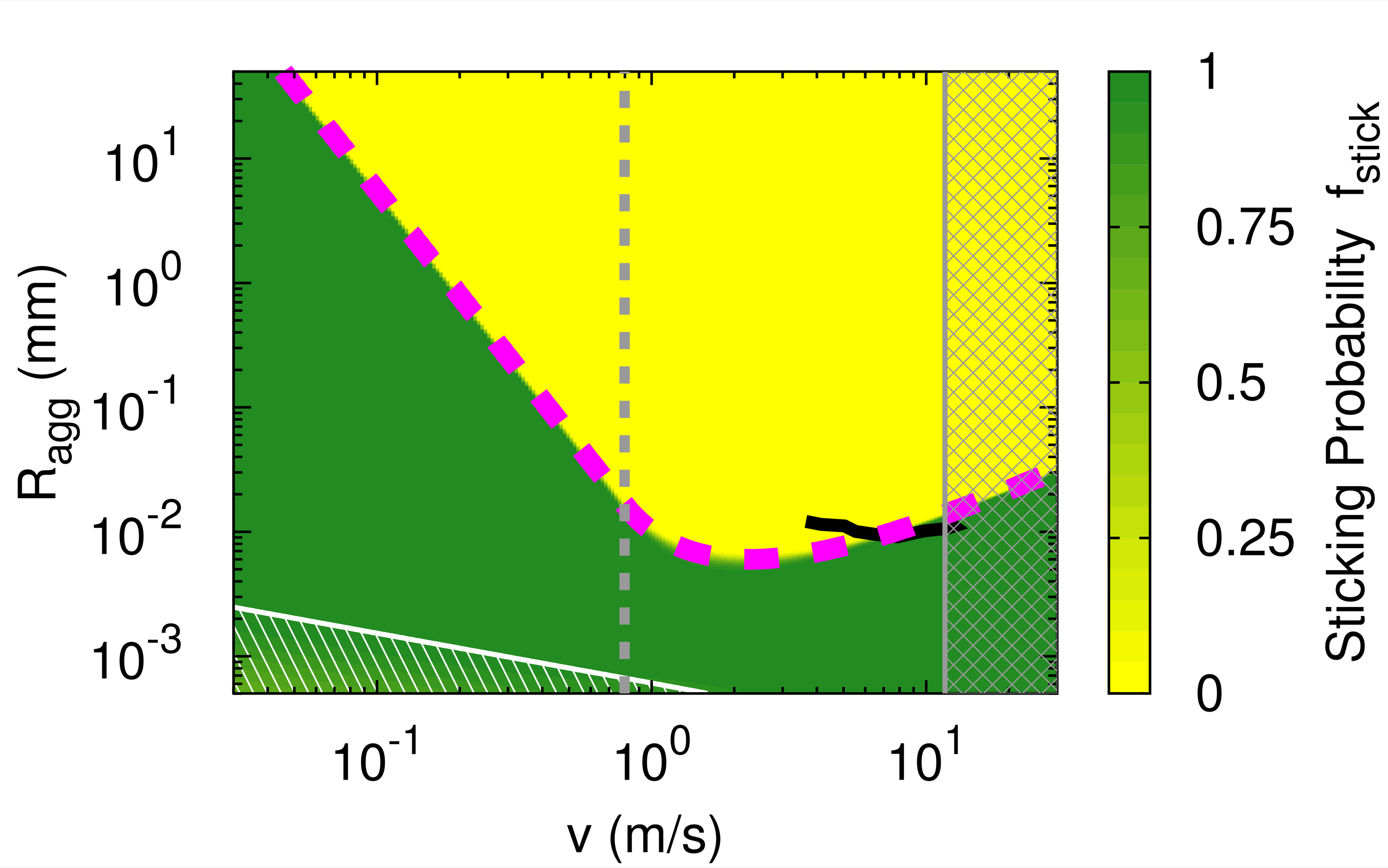}{\columnwidth}{(a) $\phi = 0.4$, $\alpha = 3$, and $\beta = 0.5$}
          \fig{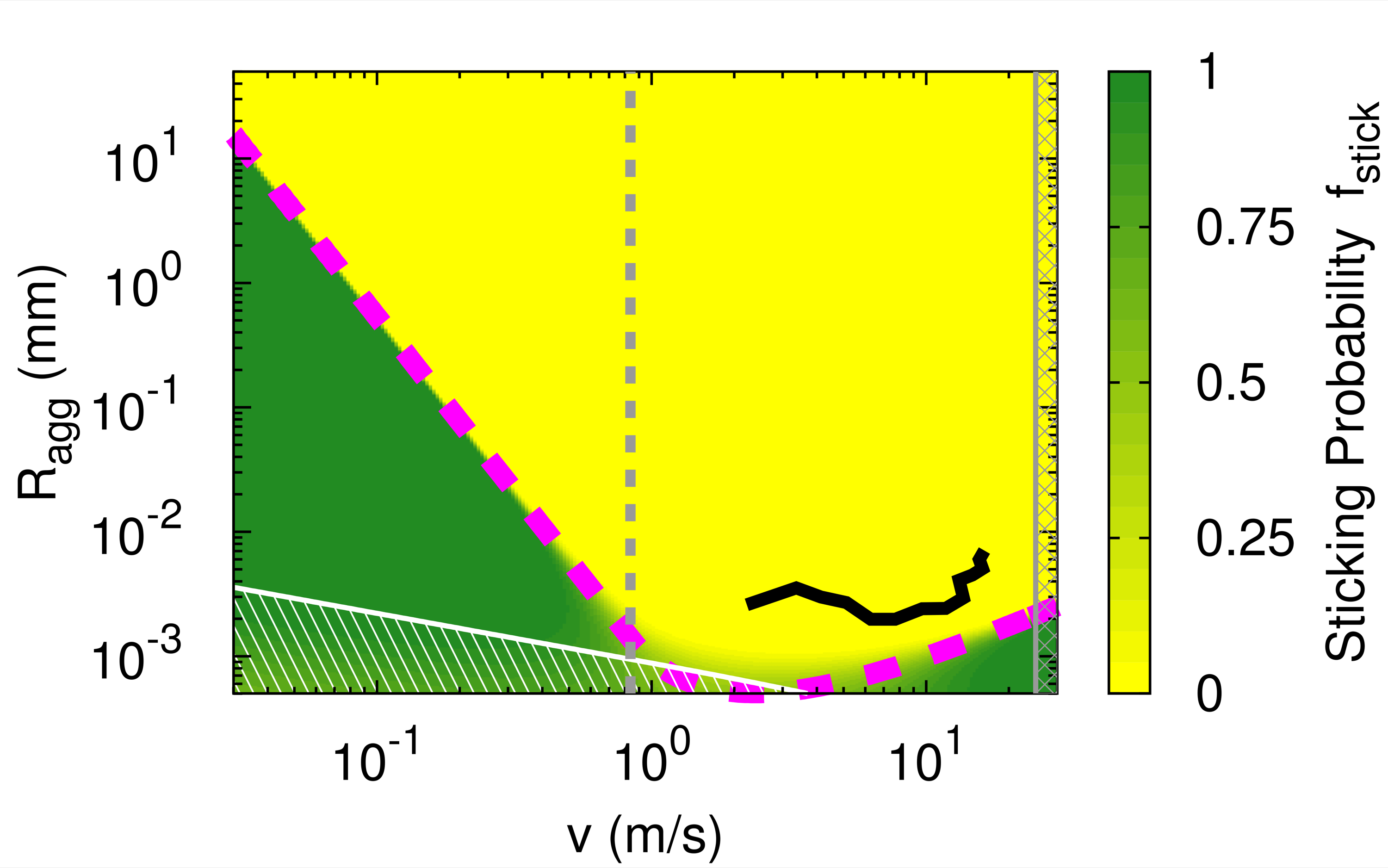}{\columnwidth}{(b) $\phi = 0.5$, $\alpha = 3$, and $\beta = 0.5$}}
\gridline{\fig{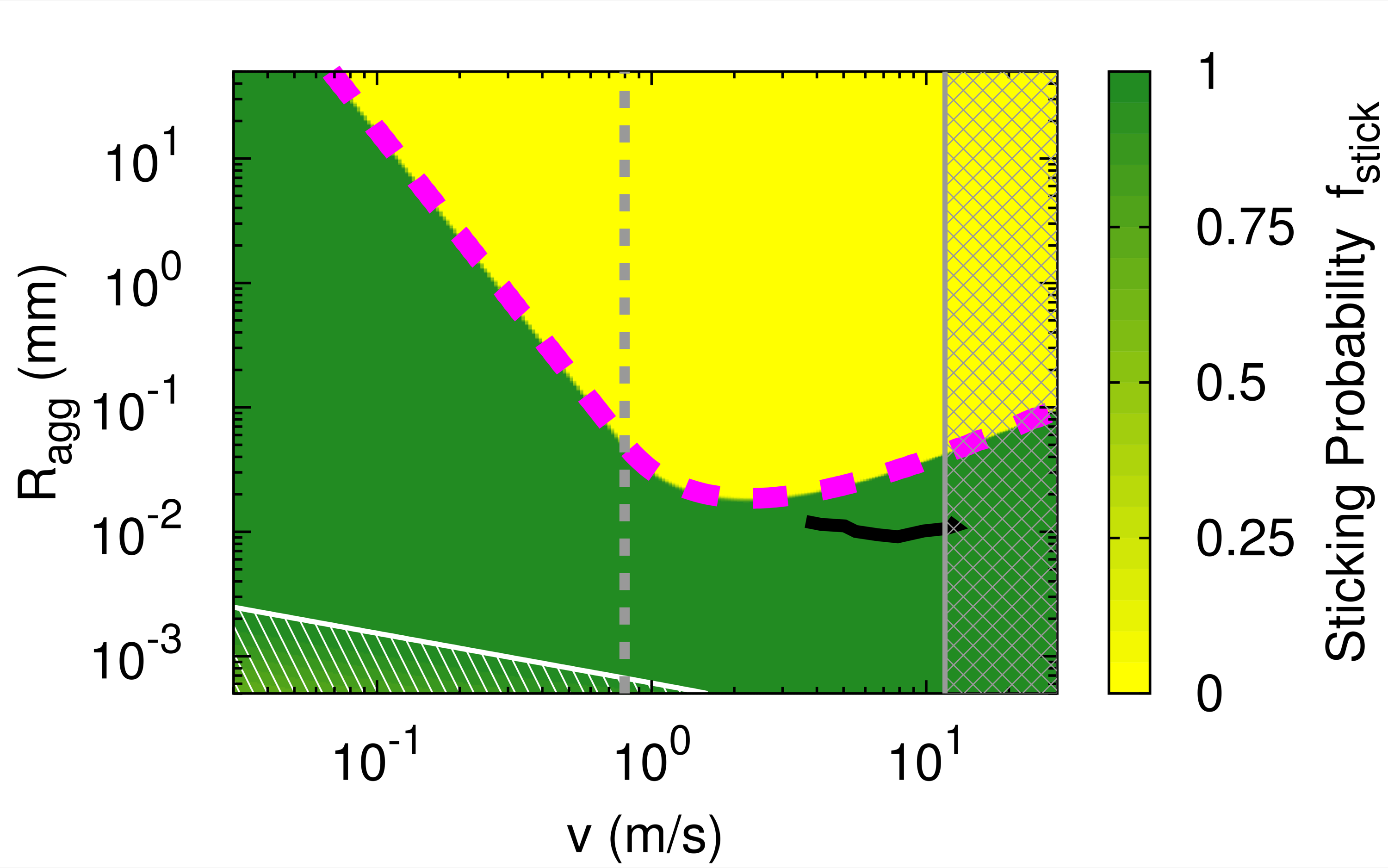}{\columnwidth}{(c) $\phi = 0.4$, $\alpha = 3$, and $\beta = 0.8$}
          \fig{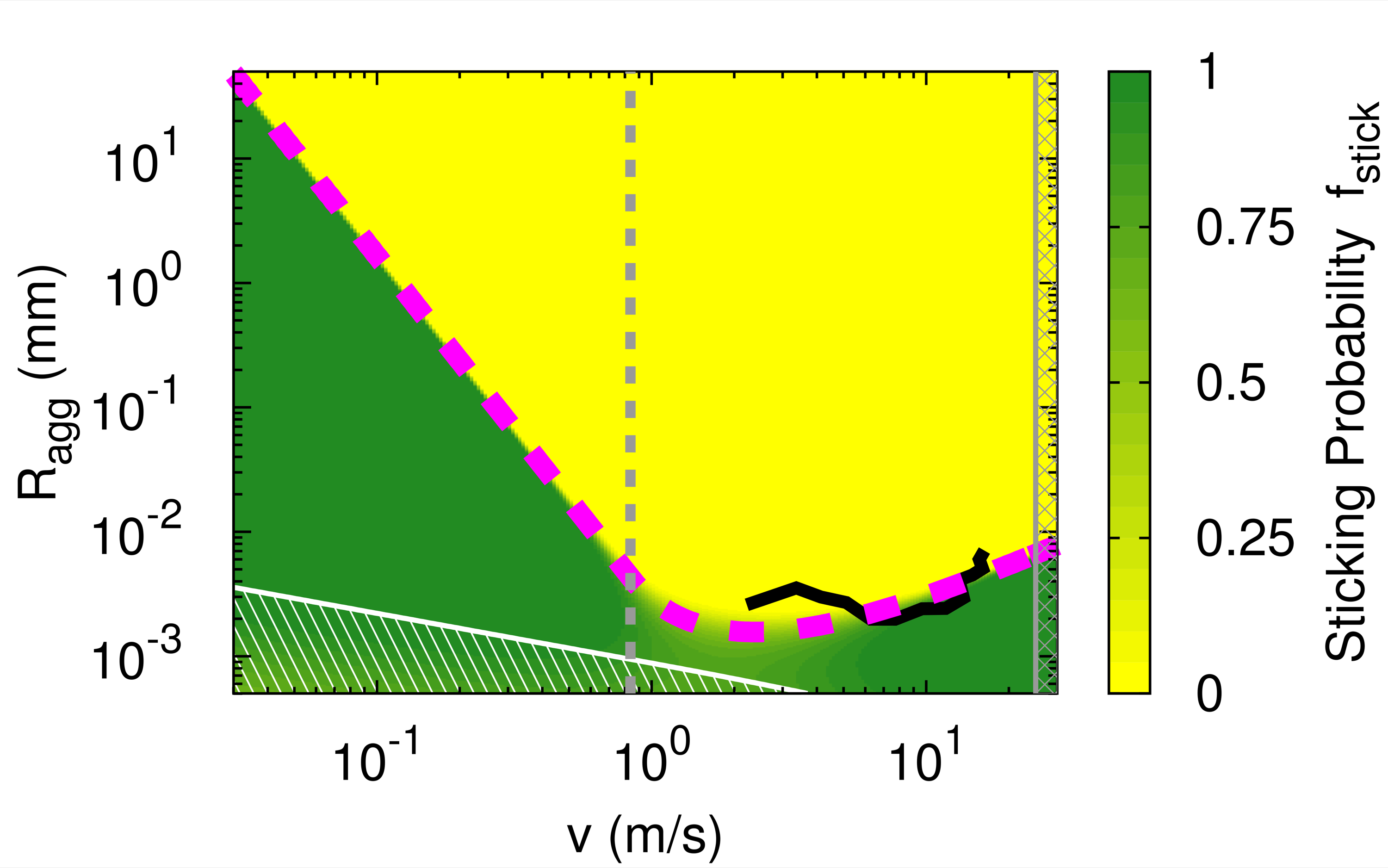}{\columnwidth}{(d) $\phi = 0.5$, $\alpha = 3$, and $\beta = 0.8$}}
\caption{
Sticking probability $f_{\rm stick}$ in the $v$--$R_{\rm agg}$ plane.
Here we fix $\alpha = 3$.
Each panel shows the result for a different set of $\phi$ and $\beta$.
The line styles and hatched regions are the same as in Figure \ref{fig:alpha}.
}
\label{fig:alpha_3}
\end{figure*}

Similarly, for $\alpha = 3$, no single value of $\beta$ can simultaneously reproduce the numerical results of \citet{2025ApJ...983...75O} for both $\phi = 0.4$ and $0.5$.
Figure \ref{fig:alpha_3} shows the $\beta$ dependence of $f_{\rm stick}$ for $\alpha = 3$ at $\phi = 0.4$ and $0.5$.
When $\alpha = 3$ and $\beta = 0.5$, the $R_{50}$ predicted by our model agrees well with the numerical results of \citet{2025ApJ...983...75O} for $\phi = 0.4$ (Figure \ref{fig:alpha_3}(a)).
However, for $\phi = 0.5$, the predicted $R_{50}$ is smaller than the boundary obtained from their numerical simulations (Figure \ref{fig:alpha_3}(b)).
In contrast, when $\beta = 0.8$, the predicted $R_{50}$ agrees well with the numerical results for $\phi = 0.5$ (Figure \ref{fig:alpha_3}(d)), whereas it becomes larger than the boundary obtained from the simulations for $\phi = 0.4$ (Figure \ref{fig:alpha_3}(c)).

\bibliography{sample701}{}
\bibliographystyle{aasjournalv7}



\end{document}